\documentclass[twocolumn,tighten,times]{aastex62}
\usepackage{amssymb}
\usepackage{amsmath}
\usepackage{amsfonts}
\usepackage{float}
\usepackage{relsize}
\usepackage{verbatim}
\usepackage{color}
\usepackage{comment}
\usepackage{graphicx}
\usepackage{bm}
\usepackage{enumitem}

\usepackage{silence}
\newcommand{\todoblank}[1]{}

\usepackage{xspace}
\graphicspath{{figures/}}

\bibpunct[; ]{(}{)}{;}{a}{}{,} %fix for the Lam et al reference with words after

\def\be{\begin{equation}}
\def\ee{\end{equation}}
\def\ba{\begin{eqnarray}}
\def\ea{\end{eqnarray}}

\newcommand{\SN}{\ensuremath{\mathrm{S/N}}} %\xspace
\newcommand{\DM}{\ensuremath{\mathrm{DM}}} %\xspace
\newcommand{\Weff}{\ensuremath{W\mathrm{_{eff}}}\xspace}
\newcommand{\Weffref}{\ensuremath{W\mathrm{_{eff,1500}}}\xspace}
\newcommand{\niss}{\ensuremath{{n\mathrm{_{ISS}}}}\xspace}
\newcommand{\DISS}{\mathrm{DISS}}
\newcommand{\R}{\mathcal{R}}
\newcommand{\Nphi}{{N_\phi}}

\newcommand{\J}{\mathrm{J}}
\newcommand{\Dtd}{\Delta t_{\mathrm{d}}}
\newcommand{\Dnud}{\Delta \nu_{\mathrm{d}}}
\newcommand{\taud}{\tau_{\mathrm{d}}}
\newcommand{\Dtdo}{\Delta t_{\mathrm{d,0}}}
\newcommand{\Dnudo}{\Delta \nu_{\mathrm{d,0}}}
\newcommand{\taudo}{\tau_{\mathrm{d,0}}}
\newcommand{\Dtdref}{\Delta t_{\mathrm{d,1500}}}
\newcommand{\Dnudref}{\Delta \nu_{\mathrm{d,1500}}}
\newcommand{\taudref}{\tau_{\mathrm{d,1500}}}
\newcommand{\nuo}{\nu_{\mathrm{0}}}

\newcommand{\Np}{N_\mathrm{p}}
\newcommand{\kJ}{k_\mathrm{J}}
\newcommand{\kJref}{k_{\J;1500}}
\newcommand{\kJo}{k_{\J;1000}}
\newcommand{\kJhatref}{\widehat{k}_{\J;1500}}

\newcommand{\kmax}{{k_{\mathrm{max}}}}

\newcommand{\Like}{\ensuremath{\mathcal{L}}}
\newcommand{\thetavec}{\bm{\theta}}
\newcommand{\chivec}{\bm{\chi}}

\newcommand{\pypulse}{\textsc{PyPulse}\xspace}
\newcommand{\nanopipe}{\textsc{nanopipe}\xspace}
\newcommand{\psrchive}{\textsc{PSRCHIVE}\xspace}

\newcommand{\tablespace}{\vspace{-0.2cm}}

\shorttitle{Frequency Dependence of Pulse Jitter}
\shortauthors{Lam et al.}

\begin{document}

\title{
The NANOGrav 12.5-Year Data Set: The Frequency Dependence of Pulse Jitter in Precision Millisecond Pulsars
}

\author[0000-0003-0721-651X]{M.\,T.\,Lam}
%\altaffiliation{NANOGrav Physics Frontiers Center Postdoctoral Fellow}
\affiliation{Department of Physics and Astronomy, West Virginia University, P.O. Box 6315, Morgantown, WV 26506, USA}
\affiliation{Center for Gravitational Waves and Cosmology, West Virginia University, Chestnut Ridge Research Building, Morgantown, WV 26505, USA}
\author[0000-0001-7697-7422]{M.\,A.\,McLaughlin}
\affiliation{Department of Physics and Astronomy, West Virginia University, P.O. Box 6315, Morgantown, WV 26506, USA}
\affiliation{Center for Gravitational Waves and Cosmology, West Virginia University, Chestnut Ridge Research Building, Morgantown, WV 26505, USA}
\author{Z.\,Arzoumanian}
\affiliation{Center for Research and Exploration in Space Science and Technology and X-Ray Astrophysics Laboratory, NASA Goddard Space Flight Center, Code 662, Greenbelt, MD 20771, USA}
\author{H.\,Blumer}
\affiliation{Department of Physics and Astronomy, West Virginia University, P.O. Box 6315, Morgantown, WV 26506, USA}
\affiliation{Center for Gravitational Waves and Cosmology, West Virginia University, Chestnut Ridge Research Building, Morgantown, WV 26505, USA}
\author{P.\,R.\,Brook}
\affiliation{Department of Physics and Astronomy, West Virginia University, P.O. Box 6315, Morgantown, WV 26506, USA}
\affiliation{Center for Gravitational Waves and Cosmology, West Virginia University, Chestnut Ridge Research Building, Morgantown, WV 26505, USA}
\author{H.\,T.\,Cromartie}
\affiliation{University of Virginia, Department of Astronomy, P.O. Box 400325, Charlottesville, VA 22904, USA}
\author{P.\,B.\,Demorest}
\affiliation{National Radio Astronomy Observatory, 1003 Lopezville Rd., Socorro, NM 87801, USA}
\author{M.\,E.\,DeCesar}
%\altaffiliation{NANOGrav Physics Frontiers Center Postdoctoral Fellow}
\affiliation{Department of Physics, Lafayette College, Easton, PA 18042, USA}
\author[0000-0001-8885-6388]{T.\,Dolch}
\affiliation{Department of Physics, Hillsdale College, 33 E. College Street, Hillsdale, Michigan 49242, USA}
\author{J.\,A.\,Ellis}
%\altaffiliation{NANOGrav Physics Frontiers Center Postdoctoral Fellow}
%\affiliation{Department of Physics and Astronomy, West Virginia University, P.O. Box 6315, Morgantown, WV 26506, USA}
%\affiliation{Center for Gravitational Waves and Cosmology, West Virginia University, Chestnut Ridge Research Building, Morgantown, WV 26505, USA}
\affiliation{Infinia ML, 202 Rigsbee Avenue, Durham NC, 27701}
\author{R.\,D.\,Ferdman}
\affiliation{School of Chemistry, University of East Anglia, Norwich, NR4 7TJ, United Kingdom}
\author{E.\,C.\,Ferrara}
\affiliation{NASA Goddard Space Flight Center, Greenbelt, MD 20771, USA}
\author[0000-0001-8384-5049]{E.\,Fonseca}
\affiliation{Department of Physics, McGill University, 3600  University St., Montreal, QC H3A 2T8, Canada}
\author{N.\,Garver-Daniels}
\affiliation{Department of Physics and Astronomy, West Virginia University, P.O. Box 6315, Morgantown, WV 26506, USA}
\affiliation{Center for Gravitational Waves and Cosmology, West Virginia University, Chestnut Ridge Research Building, Morgantown, WV 26505, USA}
\author{P.\,A.\,Gentile}
\affiliation{Department of Physics and Astronomy, West Virginia University, P.O. Box 6315, Morgantown, WV 26506, USA}
\affiliation{Center for Gravitational Waves and Cosmology, West Virginia University, Chestnut Ridge Research Building, Morgantown, WV 26505, USA}
\author{M.\,L.\,Jones}
\affiliation{Department of Physics and Astronomy, West Virginia University, P.O. Box 6315, Morgantown, WV 26506, USA}
\affiliation{Center for Gravitational Waves and Cosmology, West Virginia University, Chestnut Ridge Research Building, Morgantown, WV 26505, USA}
\author{D.\,R.\,Lorimer}
\affiliation{Department of Physics and Astronomy, West Virginia University, P.O. Box 6315, Morgantown, WV 26506, USA}
\affiliation{Center for Gravitational Waves and Cosmology, West Virginia University, Chestnut Ridge Research Building, Morgantown, WV 26505, USA}
\author{R.\,S.\,Lynch}
\affiliation{Green Bank Observatory, P.O. Box 2, Green Bank, WV 24944, USA}
\author[0000-0002-3616-5160]{C.\,Ng}
\affiliation{Department of Physics and Astronomy, University of British Columbia, 6224 Agricultural Road, Vancouver, BC V6T 1Z1, Canada}
\affiliation{Dunlap Institute, University of Toronto, 50 St. George St., Toronto, ON M5S 3H4, Canada}
\author[0000-0002-6709-2566]{D.\,J.\,Nice}
\affiliation{Department of Physics, Lafayette College, Easton, PA 18042, USA}
\author[0000-0001-5465-2889]{T.\,T.\,Pennucci}
%\altaffiliation{NANOGrav Physics Frontiers Center Postdoctoral Fellow}
\affiliation{Hungarian Academy of Sciences MTA-ELTE ``Extragalatic Astrophysics Research Group'', Institute of Physics, E\"{o}tv\"{o}s Lor\'{a}nd University, P\'{a}zm\'{a}ny P. s. 1/A, 1117 Budapest, Hungary}
\author[0000-0001-5799-9714]{S.\,M.\,Ransom}
\affiliation{National Radio Astronomy Observatory, 520 Edgemont Road, Charlottesville, VA 22903, USA}
\author[0000-0002-6730-3298]{R.\,Spiewak}
\affiliation{Centre for Astrophysics and Supercomputing, Swinburne University of Technology, P.O. Box 218, Hawthorn, Victoria 3122, Australia}
\author[0000-0001-9784-8670]{I.\,H.\,Stairs}
\affiliation{Department of Physics and Astronomy, University of British Columbia, 6224 Agricultural Road, Vancouver, BC V6T 1Z1, Canada}
%\author{D.\,R.\,Stinebring}
%\affiliation{Department of Physics and Astronomy, Oberlin College, Oberlin, OH 44074, USA}
\author{K.\,Stovall}
%\altaffiliation{NANOGrav Physics Frontiers Center Postdoctoral Fellow}
\affiliation{National Radio Astronomy Observatory, 1003 Lopezville Rd., Socorro, NM 87801, USA}
\author{J.\,K.\,Swiggum}
%\altaffiliation{NANOGrav Physics Frontiers Center Postdoctoral Fellow}
\affiliation{Center for Gravitation, Cosmology and Astrophysics, Department of Physics, University of Wisconsin-Milwaukee, P.O. Box 413, Milwaukee, WI 53201, USA}
\author{S.\,J.\,Vigeland}
%\altaffiliation{NANOGrav Physics Frontiers Center Postdoctoral Fellow}
\affiliation{Center for Gravitation, Cosmology and Astrophysics, Department of Physics, University of Wisconsin-Milwaukee, P.O. Box 413, Milwaukee, WI 53201, USA}
\author{W.\,W.\,Zhu}
\affiliation{National Astronomical Observatories, Chinese Academy of Science, 20A Datun Road, Chaoyang District, Beijing 100012, China}
%\affiliation{Max Planck Institute for Radio Astronomy, Auf dem H\"{u}gel 69, D-53121 Bonn, Germany}

%\author{G.\,Grillo}
%\affiliation{Department of Astronomy, Cornell University, Ithaca, NY 14853, USA}

%\author{J.\,S.\,Hazboun}
%\altaffiliation{NANOGrav Physics Frontiers Center Postdoctoral Fellow}
%\affiliation{Center for Advanced Radio Astronomy, University of Texas Rio Grande Valley, 1 West University Dr., Brownsville, TX 78520}
%\affiliation{Division of Physical Sciences, School of STEM, University of Washington Bothell, Box 358500, 18115 Campus Way NE, Bothell, WA 98011-8246}

%\author{J.\,E.\,Turner}
%\affiliation{Center for Gravitation, Cosmology and Astrophysics, Department of Physics, University of Wisconsin-Milwaukee, P.O. Box 413, Milwaukee, WI 53201, USA}
%\author{S.\,Chatterjee}
%\affiliation{Department of Astronomy, Cornell University, Ithaca, NY 14853, USA}
%\author{J.\,M.\,Cordes}
%\affiliation{Department of Astronomy, Cornell University, Ithaca, NY 14853, USA}
%\author{T.\,J.\,W.\,Lazio}
%\affiliation{Jet Propulsion Laboratory, California Institute of Technology, 4800 Oak Grove Drive, Pasadena, CA 91109, USA}

\correspondingauthor{M.\,T.\,Lam}
\email{michael.lam@mail.wvu.edu}

\begin{abstract}

Low-frequency gravitational-wave experiments require the highest timing precision from an array of the most stable millisecond pulsars. Several known sources of noise on short timescales in single radio-pulsar observations are well described by a simple model of three components: template-fitting from a finite signal-to-noise ratio, pulse phase/amplitude jitter from single-pulse stochasticity, and scintillation errors from short-timescale interstellar scattering variations. Currently template-fitting errors dominate, but as radio telescopes push towards higher signal-to-noise ratios, jitter becomes the next dominant term for most millisecond pulsars. Understanding the statistics of jitter becomes crucial for properly characterizing arrival-time uncertainties. We characterize the radio-frequency dependence of jitter using data on 48 pulsars in the North American Nanohertz Observatory for Gravitational Waves (NANOGrav) timing program. We detect significant jitter in 43 of the pulsars and test several functional forms for its frequency dependence; we find significant frequency dependence for 30 pulsars. We find moderate correlations of rms jitter with pulse width ($R = 0.62$) and number of profile components ($R = 0.40$); the single-pulse rms jitter is typically $\approx$1\% of pulse phase. The average frequency dependence for all pulsars using a power-law model has index $-0.42$. We investigate the jitter variations for the interpulse of PSR~B1937+21 and find no significant deviations from the main pulse rms jitter. We also test the time-variation of jitter in two pulsars and find that systematics likely bias the results for high-precision pulsars. Pulsar timing array analyses must properly model jitter as a significant component of the noise within the detector.

%PSRs~J1713+0747 and J1909$-$3744, finding no variation for PSR~J1713+0747 but some in the estimates for PSR~J1909$-$3744 likely due to other systematics such as radio-frequency interference given the high precision of the pulsar

%Previous work has looked at the broad frequency-dependence across frequency bands

\end{abstract}


\section{Introduction}
\label{sec:introduction}

Observations of recycled millisecond pulsars (MSPs) have been used for some of the most stringent tests of fundamental physics \citep{Kramer2004,Cordes+2004}, including the exploration of nuclear and plasma physics at extreme densities \citep{Demorest+2010,Antoniadis+2013,Haskell+2018}, tests of general relativity and alternate theories of gravity \citep{Will2014,Zhu+2015}, and detection and characterization of low-frequency gravitational waves \citep[GWs; e.g.,][]{EPTACW,NG11GWB}. MSPs have also been used as probes of turbulence and structures in the interstellar medium \citep[ISM;][]{Coles+2015,NG9DM,secondISMevent}, magnetic field structure in the Galaxy \citep{Han+2006,Han+2018,Gentile+2018}, kinematics in globular clusters \citep{Prager+2017}, and more \citep[e.g.,][]{Hobbs2013}.

The power of pulsar timing in testing fundamental physics comes from the highly stable rotation of MSPs and the ability to use them as precise astrophysical clocks \citep{Verbiest+2009}. Obtaining the highest timing precision possible is necessary to produce the most constraining results for experiments \citep{Cordes2013}. The precision of pulse times-of-arrival (TOAs) is limited by many noise contributions introduced along the entire propagation path, from emission at the pulsar to propagation through the ISM to measurement at radio telescopes \citep{cs2010,Verbiest+2018}. Various observational strategies can be employed to improve MSP timing precision \citep{Lee+2012,optimalfreq,optimalobs}.

%MSPs vary in their timing stability and precision, the latter of which can be improved through various observational strategies.

The North American Nanohertz Observatory for Gravitational Waves (NANOGrav; \citealt{McLaughlin2013}) observes an array of precision MSPs distributed across the sky for the detection and characterization of low-frequency (nanohertz-to-microhertz) GWs. These pulsar timing array (PTA) observations have provided robust limits on the amplitude of a stochastic GW background from a population of supermassive black hole binary (SMBHB) mergers, cosmic strings, or relic GWs from inflation \citep{NG11GWB}. We have also placed limits from single inspiraling SMBHBs \citep{NG5CW} and from the GW bursts from the merger events themselves \citep{NG5BWM}. All of these limits are made possible by the construction and continuous improvements of our PTA GW detector \citep[][]{NG11yr}. Since every Earth-pulsar baseline in the detector is different, we require a full and independent noise characterization of every arm for optimal operation \citep{NG9EN}.

Characterization of GW source properties requires the highest timing precision possible. There are many contributions to pulse TOA uncertainties, all of which have different dependencies on radio frequency \citep{optimalfreq}. The three white-noise (uncorrelated in time) TOA uncertainty contributions are the template-fitting error $\sigma_{\SN}$ from finite signal-to-noise ratio (S/N), pulse phase and amplitude jitter $\sigma_{\J}$, and the finite-scintle effect due to diffractive interstellar scintillation $\sigma_{\DISS}$ \citep[DISS;][]{NG9WN}. Jitter is caused by variations in pulse shape at the single-pulse level and in the era of high-S/N observations is becoming the next (after template-fitting errors) most important contribution to the white noise except for pulsars with high dispersion measures (DMs; the integrated line-of-sight electron density) where DISS errors will dominate due to increased scattering of the radio emission \citep{Global1713,NG9WN}. Understanding jitter is important for proper noise modeling in searches for GWs and so we must understand the statistical effects on our timing data.

We previously made measurements of jitter for MSPs in the NANOGrav 9-year data set \citep[][hereafter NG9WN for ``NANOGrav 9-year data set white noise'' paper]{NG9WN}. For each pulsar, values were estimated over a wide frequency range (typically two bands per pulsar). Similar work was performed in \citet{sod+2014}, in which per-band measurements were made and a frequency decorrelation scale was estimated.  In this work, we expand upon the previous formalism in NG9WN to model the radio-frequency dependence of jitter more broadly and quantify the role of jitter in TOA uncertainties for the increased number of pulsars in the upcoming NANOGrav 12.5-year data set. 

In \S\ref{sec:model}, we describe our methodology to model the frequency dependence of jitter. The pulsar observations and subsequent data reduction are discussed in \S\ref{sec:observations}. We summarize our analysis of jitter and both the broad and specific results in \S\ref{sec:analysis}. Finally, we discuss implications for PTAs in \S\ref{sec:discussion}.

\section{Model for Frequency-Dependent Jitter}
\label{sec:model}

We base the analysis in this paper on the methodology described in NG9WN with several modifications. Here we will discuss the new analysis methods for estimating the frequency dependence of jitter. Since jitter is independent of S/N, we aim to separate the components of the white noise that are S/N-dependent and those that are not.  In this work, we assume perfect polarization calibration and no contamination from radio-frequency interference (RFI); these will increase the variance we measure in our timing residuals, the arrival times minus those predicted by our timing model.

\subsection{Components to the TOA Uncertainty}

First we will describe the three components of the short-term timing variance. The starting point is the template-fitting procedure based on matched filtering. Matched filtering relies on the assumption that the data are described by a scaled and shifted version of an exact profile (the template) plus additive noise. In this case, the TOA uncertainty depends on the S/N and the frequency-dependent effective pulse width $\Weff(\nu)$ as \citep{cs2010}\footnote{Their form of $\Weff$ is different but similarly derived from \citet{dr1983} and the rms error is the same.}
\be
\sigma_{\SN}(S|\Weff(\nu)) = \frac{\Weff(\nu)}{S \sqrt{\Nphi}},
\ee
where $\Nphi$ is the number of bins across pulse phase $\phi$ and we write the peak-to-offpulse S/N as $S$ in equations for clarity. The effective width depends on the pulse period $P$ and the template shape $U(\phi)$ as \citep{dr1983}
\be
%\Weff = \frac{P}{\Nphi^{1/2}\left[\mathlarger{\sum}\limits_{t=1}^{\Nphi-1} \left(U_t - U_{t-1}\right)^2\right]^{1/2}}\
\Weff = \frac{P}{\Nphi^{1/2}\left[\mathlarger{\sum}\limits_{i=1}^{\Nphi-1} \left[U(\phi_i) - U(\phi_{i-1})\right]^2\right]^{1/2}}.
\label{eq:Weff}
\ee

Since individual pulse shapes vary stochastically, a data profile comprised of the average of many pulses will have a shape that differs from the template, breaking the template-fitting assumption above (\citealt{cd1985}; NG9WN). We must include an additional TOA uncertainty, one which does not depend on pulse S/N and only on the stochasticity of the single-pulse shapes. As in NG9WN, we define the dimensionless jitter parameter $\kJ \equiv \sigma_{\J,1}/P$, the ratio of the jitter-induced root-mean-square (rms) TOA variation of individual pulses to the pulse period. While individual pulse components may have different jitter statistics, we only consider the combined effect in a total rms value. The jitter component of the TOA uncertainty for an average pulse profile comprised of $\Np$ number of pulses as a function of frequency then is simply
\be 
\sigma_\J(\nu) = \frac{\sigma_{\J,1}(\nu)}{\sqrt{\Np}} = \frac{\kJ(\nu) P}{\sqrt{\Np}}.
\label{eq:sigmaJ}
\ee

Lastly, stochastic variations in the shape due to temporal pulse-broadening function (PBF), which describes random changes from interstellar scattering, also causes the data profile shape to differ from the template much like jitter \citep{cwd+1990}. The imperfect knowledge of the PBF comes from a finite number of observed scintles, intensity maxima in the time-frequency plane (dynamic spectra) with characteristic timescale $\Dtd$ and bandwidth $\Dnud$. Again, this contribution to the TOA uncertainty only depends on the stochasticity of pulse-shape variations and not pulse S/N. The scintillation bandwidth $\Dnud$ is inversely proportional to the scattering timescale of the PBF, $\taud$, as 
\be 
\Dnud = \frac{C_1}{2\pi\taud},
\label{eq:C1}
\ee
where $C_1$ is a constant of order unity that equals 1.16 for a uniform, thick Kolmogorov medium \citep{cr98}. For an observation of length $T$ and total bandwidth $B$, the number of scintles as a function of frequency $\niss(\nu)$ is 
\be 
\niss(\nu) = \left(1+\eta_t \frac{T}{\Dtd(\nu)}\right)\left(1+\eta_\nu \frac{B}{\Dnud(\nu)}\right)
\label{eq:niss}
\ee
where $\eta_t, \eta_\nu$ are the dynamic spectra filling factors $\approx 0.2$ \citep{cs2010,Levin+2016}. For a Kolmogorov medium, the scintillation timescale and bandwidth scale with frequency as \citep{cr98,NE2001}
\ba 
\Dtd(\nu) & = & \Dtdo\left(\frac{\nu}{\nuo}\right)^{6/5} \label{eq:Dtd} \\
\Dnud(\nu) & = & \Dnudo\left(\frac{\nu}{\nuo}\right)^{22/5} \label{eq:Dnud},
\ea
where $\nuo$ is the reference frequency and the subscript `0' on the scintillation parameters refers to the measurements referenced to $\nuo$. Since $\Dtd$ and $\Dnud$ vary with frequency, the number of scintles $\niss$ will also change as a function of frequency.

The scintillation noise component of the TOA uncertainty can be written as \citep{cwd+1990}
\be 
\sigma_\DISS(\nu|\Dtdo,\Dnudo) \approx \frac{\taud(\nu|\Dnudo)}{\sqrt{\niss(\nu|\Dtdo,\Dnudo)}}
\label{eq:sigmaDISS}
\ee
In practice, for most millisecond pulsars we typically measure $\Dnud$ from the dynamic spectra rather than $\taud$ from pulse broadening due to the short timescales $\taud$ involved (except at very low frequencies or very high DMs where pulse broadening becomes significant; \citealt{Levin+2016}). Therefore, we use the relations in Eqs.~\ref{eq:C1} and \ref{eq:Dnud} and write $\Dnudo$ as the input in Eq.~\ref{eq:sigmaDISS} to $\taud$ rather than $\taudo$. We note that systematic variations of $\taud$ over long timescales (months to years) result in offsets in the arrival times \citep{Palliyaguru+2015} as well as pulse shape changes \citep{Lentati+2017} but we do not include those effects in our analysis as we are only concerned with white-noise statistics of order the length of our individual observations or less.

\subsection{Modeling the Timing Residuals}

Since the timing residuals are assumed to be Gaussian distributed, the probability density function (PDF) of the residuals $\R$ is
\be 
f_{\R|S}(\R|S,\sigma_\R) = \frac{1}{\sqrt{2\pi\sigma_\R(S)^2}} e^{-\R^2/(2\sigma_\R(S)^2)}
\label{eq:gaussian}
\ee
where $\sigma_\R^2(S) = \sigma_{\SN}^2(S) + \sigma_\J^2 + \sigma_{\DISS}^2$ is a function of $S$ in only one of the three terms, i.e., the other two terms are constant in S/N. The PDF of a residual $\R$ with S/N $S$ observed at frequency $\nu$ given our model parameters is given by
\ba
\hspace{-5ex} & & f_{\R,S,\nu}(\R,S,\nu|\sigma_\R(\nu|\thetavec,\chivec)) \equiv \nonumber \\
\hspace{-5ex}  & & f_{\R,S,\nu}(\R,S,\nu|\sigma_\SN(S|\Weff(\nu)),\sigma_\J(\nu|\thetavec),\sigma_\DISS(\nu|\Dtdo,\Dnudo)) \nonumber \\
\hspace{-5ex} & & 
\label{eq:joint}
\ea
where $\thetavec$ represents jitter model parameters and $\chivec$ for simplicity represents all of the non-jitter pulsar-dependent parameters, i.e., constants describing the ISM or observational parameters such as $\Nphi$.

While short-timescale DISS causes pulse intensity fluctuations that alter the S/N (affecting $\sigma_{\SN}$ but not $\sigma_{\DISS}$ since again the latter only depends on uncertainty in the PBF), since in this work we only care about modeling the timing residuals at a given S/N, then we do not need to take the intensity distribution into consideration. Since the PDF of the residuals themselves will be Gaussian for constant S/N (or frequency), then we can separate the PDF in Eq.~\ref{eq:joint} into two terms as
\ba
& & f_{\R,S,\nu}(\R,S,\nu|\sigma_\R(\nu|\thetavec,\chivec)) \nonumber \\
& & = f_{\R|S,\nu}(\R|S,\sigma_\R(\nu|\thetavec,\chivec)) f_{S,\nu}(S,\nu|\chivec),
\ea
where the second term does not include $\thetavec$ since jitter is S/N-independent. Therefore, for the purposes of estimating jitter, we need to consider only the first term which is exactly the Gaussian distribution in Eq.~\ref{eq:gaussian}.

Therefore, we can model the variance of the residuals with a measured pulse S/N as simply
\be 
\Like(\thetavec| \left\{S_i,\R_i,\nu_i\right\},\chivec) = \prod_i f_{\R|S,\nu}(\R_i|S_i,\nu_i, \thetavec,\chivec)% \hspace{5ex}
\label{eq:likelihood}
\ee
where $i$ labels the individual measurements. Conveniently, the likelihood function for jitter depend only on the timing data as long as the other quantities can be estimated independently.

\section{Observations and Data Reduction}
\label{sec:observations}

The NANOGrav 12.5-year data set will contain new methods for timing and noise model parameterization as well as software developments over its previous data sets. However, the data reduction to obtain processed pulse profiles after data acquisition remains largely similar to the procedures used in \citet{NG11yr}, the NANOGrav 11-year data set paper, which we will discuss here with the relevant modifications. For reference, Table~\ref{table:input} lists the pulsars, the telescopes used to observe them, and their spin periods and DMs. Other parameters listed will be discussed in the next subsections. We excluded PSR~J1747$-$4036 from our analysis because of its very low S/N and therefore it is unlisted in the table (the pulsar was similarly excluded in NG9WN).

We observed pulsars with two radio telescopes: the 305-m William E. Gordon Telescope at the Arecibo Observatory (AO) and the 100-m Green Bank Telescope at the Green Bank Observatory (GBO). While two generations of backends were used at each facility, in this work we only used data taken with the more recent Puerto Rican and Green Bank Ultimate Pulsar Processing Instruments (PUPPI and GUPPI, respectively; \citealt{drd+2008,fdr2010}) as they can process much larger bandwidths. The increased bandwidths allow NANOGrav to boost the averaged pulse S/Ns and allows for more scintillation maxima to be observed \citep{Pennucci}. 

Pulsars were observed with two or more receivers per epoch to estimate the time-varying dispersion measure. All pulsars were observed in the 1500~MHz band with a maximum of 800~MHz of bandwidth. At GBO, we observed all pulsars with the 820~MHz receiver as well, covering 200~MHz of bandwidth. At AO, either the 430~MHz ($B = 50$~MHz) or 2300~MHz ($B = 800$~MHz) receivers were also used, with the exception of PSR~J2317+1439 in which the 327~MHz ($B = 50$~MHz) receiver was used along with the 430 and 1500~MHz receivers. Two pulsars, PSRs~J1713+0737 and B1937+21, were observed with both telescopes. Unlike in NG9WN, since we do not attempt to characterize the scintillation statistics of our data set\footnote{In NG9WN, we attempted to model the intensity distribution as solely due to DISS, which fit the observed data for some pulsars well but for others not so well. Refractive intensity variations over long timescales (weeks to years) were not modeled, nor were changes in the diffractive scintillation parameters over those same timescales.}, we combined the data from both telescopes for these two pulsars as Eq.~\ref{eq:gaussian} still applies. 

The raw pulse profiles were calculated over small ($\sim1-15$~s) subintegrations depending on the receiver/backend combination; the length of a typical observation per receiver was approximately 30~minutes. These raw profiles were calculated from the average of many single pulses using an initial timing model (folding) created from previous observations that accounted for the spin, astrometric, and binary (if applicable) parameters, and were also coherently dedispersed. The profiles were split into 2048 bins that spanned pulse phase; the absolute time duration of each phase bin therefore varies between pulsars. To reduce the data volume, profiles were averaged in time, to $\sim$80~s at AO and $\sim$120~s at GBO.

To calibrate the data, we used the \psrchive package \citep{psrfits,psrchive} accessed via \nanopipe \citep{nanopipe}, with the broad procedure as follows. At each telescope, we injected and recorded a broadband noise source prior to each pulsar observation to calibrate the differential phase and gain offsets between both hands of polarization. We did not assume that the two hands of polarization in the noise source were equal. Every month we observed the noise source at each telescope and frequency band both on and off the position of a bright unpolarized quasar as a reference for absolute calibration; we assumed that the noise sources were stable over this timescale. We also used these quasars as references for flux calibration.

Slight timing mismatches in the interleaved analog-to-digital converters samplers in the GUPPI/PUPPI backends resulted in a frequency-reversed ``ghost image'' of the pulse appearing in our data (see \citealt{Kurosawa+2001} for a general discussion of the effect). After the pulse profiles are dedispersed, the image appears as a very low amplitude negatively-dispersed copy of the signal, more prominent in bright low-DM pulsars (see Figure 1.2 of \citealt{Thesis} for an example from data of PSR~J1713+0747 in \citealt{Global1713}). While faint, the timing data was noticeably offset in the residuals as shown by an outlier analysis (\citealt{outlier}, see also \citealt{NG11yr}), especially at frequencies where the image crossed the main pulse and therefore required mitigation. The amount of leakage was measured over time from the calibration data. After the image rejection parameters were measured, we used the correction algorithms (again, see \citealt{Kurosawa+2001}) implemented in \psrchive with the relevant parameters given in \nanopipe to remove the images. The full details of this image removal will be discussed in the future NANOGrav 12.5-year data set paper.

Once the polarization profiles were calibrated properly, we summed them to form the intensity profiles used in this work. To mitigate narrowband RFI, we first excised consistent known sources from our data. We then implemented an algorithm to calculate the off-pulse intensity variation across a rolling 20-frequency-channel-wide window per subintegration. If the variation was four times the median value in that window, those channels were also removed. %Finally, the profiles were averaged in time and 

In the very low-S/N limit, template fitting fails due to matches with noise features. The effect is mitigated somewhat for pulses with wide template shapes as the S/N of the cross-correlation function between the template and the data profile is larger than the S/N of the profile itself (NG9WN). However, the standard $\sigma_{\SN}$ form underestimates the TOA uncertainty and becomes non-Gaussian (see Appendix B of \citealt{NG9yr}). To increase our S/N, we averaged our profiles over a number of frequency channels, with that number varying by frequency band to avoid significant contamination from dispersive smearing in the individual channels, discussed in the next section.

\subsection{From Pulse Profiles to Short-Term Residuals}

In the recording of each subintegration of data discussed above, pulse profiles were folded and dedispersed using a pre-computed initial pulsar-timing model. Using \pypulse \citep{pypulse}, we calculated the pulse arrival phases (in time units) within each subintegration, the ``initial timing residuals'' $\delta t(\nu,t)$ (NG9WN), via a Fourier-domain method \citep{Taylor1992}. Given the stability of our pulsar ephemerides over long timescales, we assumed that any drift in the pulse arrival times within an observation would be small and could be described by low-order polynomial correction terms in phase. For example, for isolated pulsars, the typical pulse-smearing error in the initial folding period causes timing uncertainties of the order of 10s of picoseconds, though for binary pulsars the uncertainty can be of the order of 10s to 100s of nanoseconds and so we must account for these corrections (the systematic drift in the TOAs is several orders of magnitude smaller). See Appendix A of NG9WN for the discussion of these and many other effects that cause errors in the initial timing model on short timescales.

The initial timing model can be written as
\be
\delta t(\nu,t) = K(\nu) + a t + bt^2 + \cdots + n(\nu,t)
\label{eq:dt}
\ee
where $a$ and $b$ are frequency-independent coefficients describing the low-order polynomial drift correction, $n(\nu,t)$ is the additive noise composed of the three white-noise terms, and $K(\nu)$ is a constant offset per-frequency that describes all unaccounted for pulse-profile shape evolution in frequency and epoch-dependent dispersion and scattering delays. The ``short-term'' residuals are then
\be 
\R(\nu,t) \equiv \hat{n}(\nu,t) \approx \delta t(\nu,t) - \left[\widehat{K}(\nu) + \hat{a}t + \hat{b}t^2\right],
\label{eq:residuals}
\ee
where the carets denote estimated quantities. Subtraction of $\widehat{K}(\nu)$ removes all of the unknown frequency dependence between the subbands. Note that since we remove the frequency dependence of the residuals and do not fit a model to reference the arrival times to ``infinite frequency'' as is often done when attempting to combine TOAs to model interstellar propagation effects in typical timing models, we are not concerned with systematic uncertainties in the arrival times due to this referencing to infinite frequency such as from dispersive-delay removal \citep{optimalfreq}. While the fit for $a$, $b$, and $K(\nu)$ will remove some amount of variance we wish to measure from the residuals, the total amount will be small for many white-noise residuals (e.g., $\sim$16 bands $\times$ $\sim$15 sub-integrations for the 1500~MHz band) that are uncorrelated in time.

As mentioned previously, averaging of the pulses in frequency causes errors from dispersive smearing. While the raw data were coherently dedispersed, because the initial-timing model used to fold and dedisperse the data does not account for any time variations of DM over many years, the DM value used for coherent dedispersion can be significantly different from the true DM. The timing perturbation from pulse smearing over a subband (channel) bandwidth $B_{\rm chan}$ due to a variation $\delta \DM$ is \citep[][NG9WN]{Cordes2002}
\be 
\delta t_{\delta \DM} = 8.3~\mathrm{\mu s} \left(\frac{\delta\DM}{\mathrm{pc~cm^{-3}}}\right) \left(\frac{B_{\rm chan}}{\mathrm{MHz}}\right) \left(\frac{\nu}{\mathrm{GHz}}\right)^{-3}.
\label{eq:smear}
\ee
Systematic variations in DM over the length of our data set can be up to $\approx10^{-3}$~pc~cm$^{-3}$ \citep{NG9DM}; many pulsars show lower-amplitude variations overall and the variations between epochs typically are significantly smaller than $10^{-3}$~pc~cm$^{-3}$. However, the timing perturbation will be constant over the time of an observation for a specific frequency (since the offset $\Delta \DM$ will be constant) and therefore will be removed by the fit for the offset $K(\nu)$. By taking the variance of the perturbations in Eq.~\ref{eq:smear}, the timing error is given by $\sigma_{\delta t_{\delta \DM}} = 8.3~\mathrm{\mu s}~\sigma_{\delta \DM}~B_{\rm chan}~ \nu^{-3}$ in the same units; these uncertainties will increase the variance of the residuals and bias our estimates of jitter. The uncertainties on DM for our pulsars are typically much smaller, by roughly one to three orders of magnitude. For PSR~J2317+1439 observed at our lowest frequencies, where the DM errors are of order $\sigma_{\delta \DM} \sim 10^{-5}$~pc~cm$^{-3}$, the timing error in the 327~MHz band with 0.78125~MHz channels (64 over the band) is only 2~ns which is small when compared to the rms of the residuals. However, when averaging over the entire 50 MHz, the error grows to $\sigma_{\delta \DM} \approx 120$~ns. Therefore, for the 327 and 430~MHz bands, rather than average over the 50 MHz to build S/N as discussed previously, we choose to keep the full frequency resolution of the data\footnote{In NG9WN, we used 50-MHz channels for all frequency bands and therefore we did not account for this smearing error.}. For the 820, 1400, and 2300~MHz bands, we averaged the pulses into 50~MHz channels to build S/N; we have for the same DM error above of $10^{-5}$~pc~cm$^{-3}$, $\sigma_{\delta t_{\delta \DM}} \approx 7.5$~ns for 50~MHz channels centered at 820~MHz and therefore the error is not a significant component of the total rms.

% For 50-MHz channels at an observing frequency of 1.5~GHz (1500~MHz), the resulting timing perturbation is $\sim120$~ns; again, this is the maximum deviation if the DM difference is $\approx10^{-3}$~pc~cm$^{-3}$. For our lowest frequency band, 327~MHz, channels of 0.5~MHz are needed for similar perturbation amplitudes.

%In practice, for each pulsar/backend/receiver combination, we took the measured effective pulse width and fit the pulses in a small time-frequency pulse profile ($\sim$1-2 minute and 50~MHz) to get the S/N and ``TOA", then measured $\sigma_\C$ per combination. From $\sigma_\C$, using our knowledge of the scintillation parameters, for example from Levin et al. 2016, Keith et al. 2013, NE2001, etc., we then found $\sigma_\J$ simply by taking $\sigma_\J^2 = \sigma_\C^2 - \sigma_{\DISS}^2$. We found many upper limits in $\sigma_\C$ and thus $\sigma_\J$. In some cases, $\sigma_{\DISS} > \sigma_\C$, which we noted in the paper. While $\niss$ was measured directly from the PDF of $S$, since these were often messy, when estimating the $\sigma_{\DISS}$ contribution to $\sigma_\C$, we simply used the previously-estimated scintillation parameters.
\begin{deluxetable*}{lccccccc}
\tablecolumns{8}
\tabletypesize{\scriptsize}
\tablecaption{Reference and Input Pulsar Parameters\label{table:input}}
\tablehead{
\colhead{Pulsar} & \colhead{Telescope} & \colhead{Period} & \colhead{DM} & \colhead{$W_{\rm eff,1500}$} & \colhead{$\Delta t_{\rm d,1500}$} & \colhead{$\Delta \nu_{\rm d,1500}$} & \colhead{$\tau_{\rm d,1500}$}\\
\colhead{} & \colhead{} & \colhead{(ms)} & \colhead{$\left(\mathrm{pc~cm^{-3}}\right)$}  & \colhead{($\mu$s)}  & \colhead{(s)} & \colhead{(MHz)} & \colhead{(ns)}  
}
\startdata
\vspace{-1ex} J0023+0923 & AO & 3.05 & 14.32 & 417.4 & 1210$^{\rm a}$ & 20$^{\rm b}$ & 7.8\\
\vspace{-1ex} J0030+0451 & AO & 4.87 & 4.33 & 527.8 & 44300$^{\rm a}$ & 1330$^{\rm b}$ & 0.12\\
\vspace{-1ex} J0340+4130 & GBT & 3.30 & 49.59 & 489.4 & 430 & 9.1$^{\rm b}$ & 17\\
\vspace{-1ex} J0613$-$0200 & GBT & 3.06 & 38.78 & 337.1 & 4500 & 11$^{\rm b}$ & 14\\
\vspace{-1ex} J0636+5128 & GBT & 2.87 & 11.11 & 447.4 & 1170$^{\rm c}$ & 97 & 1.9$^{\rm c}$\\
\vspace{-1ex} J0645+5158 & GBT & 8.85 & 18.25 & 618.7 & 780 & 30$^{\rm b}$ & 6.1\\
\vspace{-1ex} J0740+6620 & GBT & 2.89 & 14.96 & 273.9 & 1070$^{\rm a}$ & 59$^{\rm b}$ & 3.2\\
\vspace{-1ex} J0931$-$1902 & GBT & 4.64 & 41.49 & 381.6 & 640 & 50 & 3.2\\
\vspace{-1ex} J1012+5307 & GBT & 5.26 & 9.02 & 601.7 & 1350 & 66 & 2.4\\
\vspace{-1ex} J1022+1001 & AO & 16.45 & 10.25 & 1454.2 & 1320 & 130 & 1.4\\
\vspace{-1ex} J1024$-$0719 & GBT & 5.16 & 6.48 & 570.6 & 4180 & 47$^{\rm b}$ & 3.4\\
\vspace{-1ex} J1125+7819 & GBT & 4.20 & 11.22 & 648.0 & 1530 & 120$^{\rm b}$ & 1.5\\
\vspace{-1ex} J1453+1902 & AO & 5.79 & 14.06 & 826.7 & 1420 & 61 & 3.0\\
\vspace{-1ex} J1455$-$3330 & GBT & 7.99 & 13.57 & 999.0 & 4670$^{\rm a}$ & 70$^{\rm b}$ & 2.3\\
\vspace{-1ex} J1600$-$3053 & GBT & 3.60 & 52.33 & 425.5 & 270 & $9.0 \times 10^{-2}$ & $1.8 \times 10^{3}$\\
\vspace{-1ex} J1614$-$2230 & AO & 3.15 & 34.49 & 389.3 & 480 & 9.0 & 18\\
\vspace{-1ex} J1640+2224 & AO & 3.16 & 18.46 & 453.8 & 1030$^{\rm d}$ & 56$^{\rm b}$ & 5.8\\
\vspace{-1ex} J1643$-$1224 & GBT & 4.62 & 62.30 & 975.1 & 580$^{\rm a}$ & $2.2 \times 10^{-2}$ & $7.2 \times 10^{3}$$^{\rm a}$\\
\vspace{-1ex} J1713+0747 & AO/GBT & 4.57 & 15.92 & 530.4 & 2860 & 21$^{\rm b}$ & 7.5\\
\vspace{-1ex} J1738+0333 & AO & 5.85 & 33.77 & 627.1 & 600 & 17 & 9.5\\
\vspace{-1ex} J1741+1351 & AO & 3.75 & 24.20 & 380.5 & 2350$^{\rm a}$ & 17 & 9.3$^{\rm a}$\\
\vspace{-1ex} J1744$-$1134 & GBT & 4.07 & 3.09 & 511.3 & 2070$^{\rm a}$ & 42$^{\rm b}$ & 3.8\\
\vspace{-1ex} J1832$-$0836 & AO & 2.72 & 28.19 & 186.0 & 580 & 10$^{\rm b}$ & 18\\
\vspace{-1ex} J1853+1303 & AO & 4.09 & 30.57 & 337.0 & 1460$^{\rm a}$ & 13$^{\rm b}$ & 13\\
\vspace{-1ex} B1855+09 & AO & 5.36 & 13.30 & 754.2 & 1460 & 5.2 & 31\\
\vspace{-1ex} J1903+0327 & AO & 2.15 & 297.52 & 390.0 & 12 & $1.7 \times 10^{-3}$ & $9.3 \times 10^{4}$\\
\vspace{-1ex} J1909$-$3744 & GBT & 2.95 & 10.39 & 258.6 & 2260 & 39$^{\rm b}$ & 4.1\\
\vspace{-1ex} J1910+1256 & AO & 4.98 & 38.07 & 631.6 & 1220 & 2.3 & 69$^{\rm e}$\\
\vspace{-1ex} J1911+1347 & AO & 4.63 & 30.99 & 459.9 & 1480$^{\rm a}$ & 37$^{\rm b}$ & 5.0\\
\vspace{-1ex} J1918$-$0642 & GBT & 7.65 & 26.46 & 876.7 & 800 & 15$^{\rm b}$ & 11\\
\vspace{-1ex} J1923+2515 & AO & 3.79 & 18.86 & 499.3 & 2260 & 22 & 7.4\\
\vspace{-1ex} B1937+21 & AO/GBT & 1.56 & 71.09 & 144.1 & 330 & 2.8$^{\rm b}$ & 57\\
\vspace{-1ex} J1944+0907 & AO & 5.19 & 24.34 & 916.9 & 1810 & 11$^{\rm b}$ & 15\\
\vspace{-1ex} J1946+3417 & GBT & 3.17 & 111.11 & 453.6 & 350 & 0.84$^{\rm b}$ & 220\\
\vspace{-1ex} B1953+29 & AO & 6.13 & 104.50 & 818.7 & 320 & 2.9 & 55\\
\vspace{-1ex} J2010$-$1323 & GBT & 5.22 & 22.18 & 516.1 & 240 & 0.30 & 610\\
\vspace{-1ex} J2017+0603 & AO & 2.90 & 23.92 & 235.4 & 910 & 6.9$^{\rm b}$ & 23\\
\vspace{-1ex} J2033+1734 & AO & 5.95 & 25.08 & 818.3 & 1480 & 22$^{\rm b}$ & 7.4\\
\vspace{-1ex} J2043+1711 & AO & 2.38 & 20.71 & 174.5 & 1900 & 63 & 2.9\\
\vspace{-1ex} J2145$-$0750 & GBT & 16.05 & 9.00 & 1826.7 & 2140 & 86$^{\rm b}$ & 1.8\\
\vspace{-1ex} J2214+3000 & AO & 3.12 & 22.54 & 555.2 & 3400$^{\rm a}$ & 48$^{\rm b}$ & 3.3\\
\vspace{-1ex} J2229+2643 & AO & 2.98 & 22.73 & 568.8 & 1610 & 58 & 3.2\\
\vspace{-1ex} J2234+0611 & AO & 3.58 & 10.77 & 387.1 & 1350 & 44 & 4.2\\
\vspace{-1ex} J2234+0944 & AO & 3.63 & 17.83 & 544.5 & 2180 & 240 & 0.76\\
\vspace{-1ex} J2302+4442 & GBT & 5.19 & 13.79 & 633.1 & 1250 & 23$^{\rm b}$ & 6.9\\
\vspace{-1ex} J2317+1439 & AO & 3.45 & 21.90 & 383.1 & 2740 & 9.9$^{\rm b}$ & 16\\
\vspace{-1ex} J2322+2057 & AO & 4.81 & 13.36 & 434.7 & 3580$^{\rm d}$ & 42$^{\rm b}$ & 3.8 \vspace{1ex} \\
\vspace{-2ex}\\\enddata
\tablenotetext{}{The spin period and DM were taken from NANOGrav timing data. The effective width $W_{\rm eff,1500}$ was calculated from the wideband templates. If $\Delta \nu_{\rm d}$ or $\tau_{\rm d}$ has a reference, then we estimate the other via the relationship in Eq.~\ref{eq:C1}. Otherwise, all other scintillation parameters without references were estimated from \citet{NE2001}. Effective width and scintillation measurements were referenced to 1500~MHz.\vspace{-2ex}}
\tablenotetext{a}{\vspace{-2ex}\citet{Keith+2013}.}
\tablenotetext{b}{\vspace{-2ex}\citet{Levin+2016}.}
\tablenotetext{c}{\vspace{-2ex}\citet{Nicastro+2001}.}
\tablenotetext{d}{\vspace{-2ex}\citet{Johnston+1998}.}
\tablenotetext{e}{\citet{Champion+2008}.}
\end{deluxetable*}

\subsection{Wideband Templates}

The NANOGrav 12.5-year data set will contain two sets of timing methodologies per epoch per frequency band: arrival times estimated per channel \citep[see also][]{NG11yr} and a single ``wideband'' TOA with a simultaneous DM estimate \citep[][with more specific discussion in an upcoming paper, Pennucci et al. in prep]{Pennucci+2014}. For both methods, we created smoothed average-profile shapes per-frequency-band used in the template-matching procedure to estimate the TOAs. For the former, a single template shape was produced by aligning and adding the sum of the pulse profiles followed by wavelet denoising to smooth the template; shape deviations from the template over the band \citep[see e.g.,][]{Global1713} that cause TOA offsets are corrected for in the timing model. For the latter methodology, a two-dimensional ``pulse portrait'' was created, providing information about the template shape as a function of frequency which mitigated the need for these frequency-dependent arrival-time corrections.

Since the wideband timing method produces template shapes that vary as a function of frequency, we can more finely measure the shape and effective width of the templates at a specific frequency. Therefore, we were able to estimate $\Weff(\nu)$ as in Eq.~\ref{eq:Weff}. In Table~\ref{table:input}, we provide the estimates of $\Weff(1500~\mathrm{MHz})$, which for brevity we denote as $\Weffref$.

\begin{figure*}[t!]
\centering
	\includegraphics[width=0.85\textwidth]{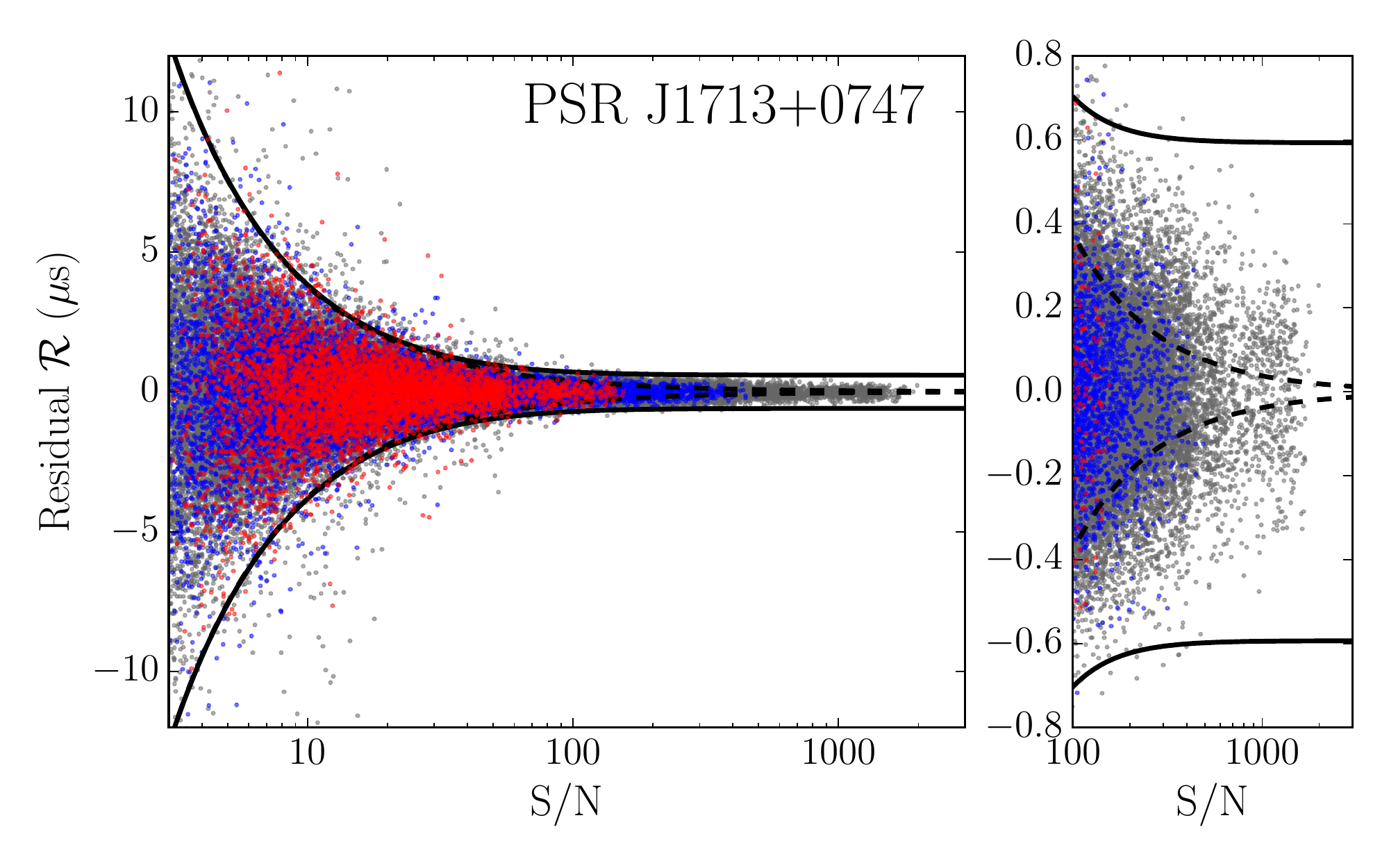}
  \caption{\footnotesize Short-term residuals $\R(\nu,t)$ vs S/N for PSR~J1713+0747. The colors show residuals grouped by frequency band for visual clarity\todoblank{not sure how best to show this clearly}: red for 820~MHz, gray for 1500~MHz, blue for 2300~MHz. The solid lines show $\pm 3\sigma_\R$ where all parameters have been scaled to 1500~MHz and the $\sigma_\J$ component is taken from the constant model (Model A, see Table~\ref{table:models}) and has been scaled to 120~s using Eq.~\ref{eq:sigmaJ} (note that many of the residuals were measured with AO and have 80~s subintegrations and so the scaling of the lines are not perfect). The contribution of the $\sigma_\DISS$ component (scaled using Eq.~\ref{eq:niss} assuming 120~s and 800~MHz bandwidth as for our 1500~MHz bands) on $\sigma_\R$ is negligible on the scale of the plot. The dashed lines show $\pm 3\sigma_\SN$, again assuming $\nu = 1500$~MHz. The right panel is a zoomed-in section of the left panel; one can see that even the approximate $\sigma_\SN(S)$ does not adequately describe the variance of the residuals.} %\sigma_\R 3\sqrt{\sigma_{\SN}(S|\Weffref)^2 + \sigma_\J^2}
  \label{fig:J1713}
\end{figure*}

\subsection{Scintilation Parameters}

We used scintillation parameters taken from the literature in order to estimate the scintillation noise component $\sigma_{\DISS}(\nu)$ per pulsar. Table~\ref{table:input} shows values of $\Dtdref$ and $\Dnudref$ (the quantities are referenced to 1500~MHz as specified by the subscript) we used to estimate $\sigma_{\DISS}(\nu|\Dtdref,\Dnudref)$ as per Eq.~\ref{eq:sigmaDISS}, as well as $\taudref$ for reference. Citations for the values are provided in the table, and were taken primarily from \citet{Keith+2013} and \citep{Levin+2016} and references therein. When values were unavailable, we estimated them using the NE2001 electron density model \citep{NE2001}. We assumed that $C_1 = 1.16$ for conversions between $\taud$ and $\Dnud$ (Eq.~\ref{eq:C1}) and $\eta_t = \eta_\nu = 0.2$ for the calculation of $\niss$ (Eq.~\ref{eq:niss}). Typically, $\sigma_{\DISS}$ is small except at the lowest frequencies and for pulsars with high DM in which the $\taud$ is large \citep{Bhat+2004}.

\section{Data Analysis and Results}
\label{sec:analysis}

For each pulse labeled $i$ observed at frequency $\nu_i$ with S/N $S_i$ and residual $\R_i$ estimated from Eq.~\ref{eq:residuals} (i.e., $\R(\nu,t)$), and using our knowledge of $\Weffref(\nu)$, $\Dtdref$, and $\Dnudref$ (all components of $\chivec$), we were able to model the PDFs for each residual using Eqs.~\ref{eq:gaussian} and \ref{eq:joint}. Using all of the data with $S>3$, we could therefore estimate the likelihood function $\Like(\hat{\thetavec}| \left\{S_i,\R_i,\nu_i\right\},\chivec)$ in Eq.~\ref{eq:likelihood} for each pulsar.

We used the Markov-chain Monte Carlo (MCMC) method via the \textsc{emcee} package \citep{emcee} to explore the likelihood space and estimate the parameters $\hat{\thetavec}$ for different models of the frequency-dependence of jitter.  We list the models and their functional forms in Table~\ref{table:models}. We referenced our rms jitter parameters to the single-pulse values, i.e., $\sigma_{\J,1}$, as that is the underlying fundamental quantity, even though the residuals were formed from profiles averaged from $\sim1-2$ minutes of data. Again, one can scale the single-pulse rms jitter values using Eq.~\ref{eq:sigmaJ}.

\begin{figure}[ht!]
\hspace{-5ex}
\includegraphics[width=0.5\textwidth]{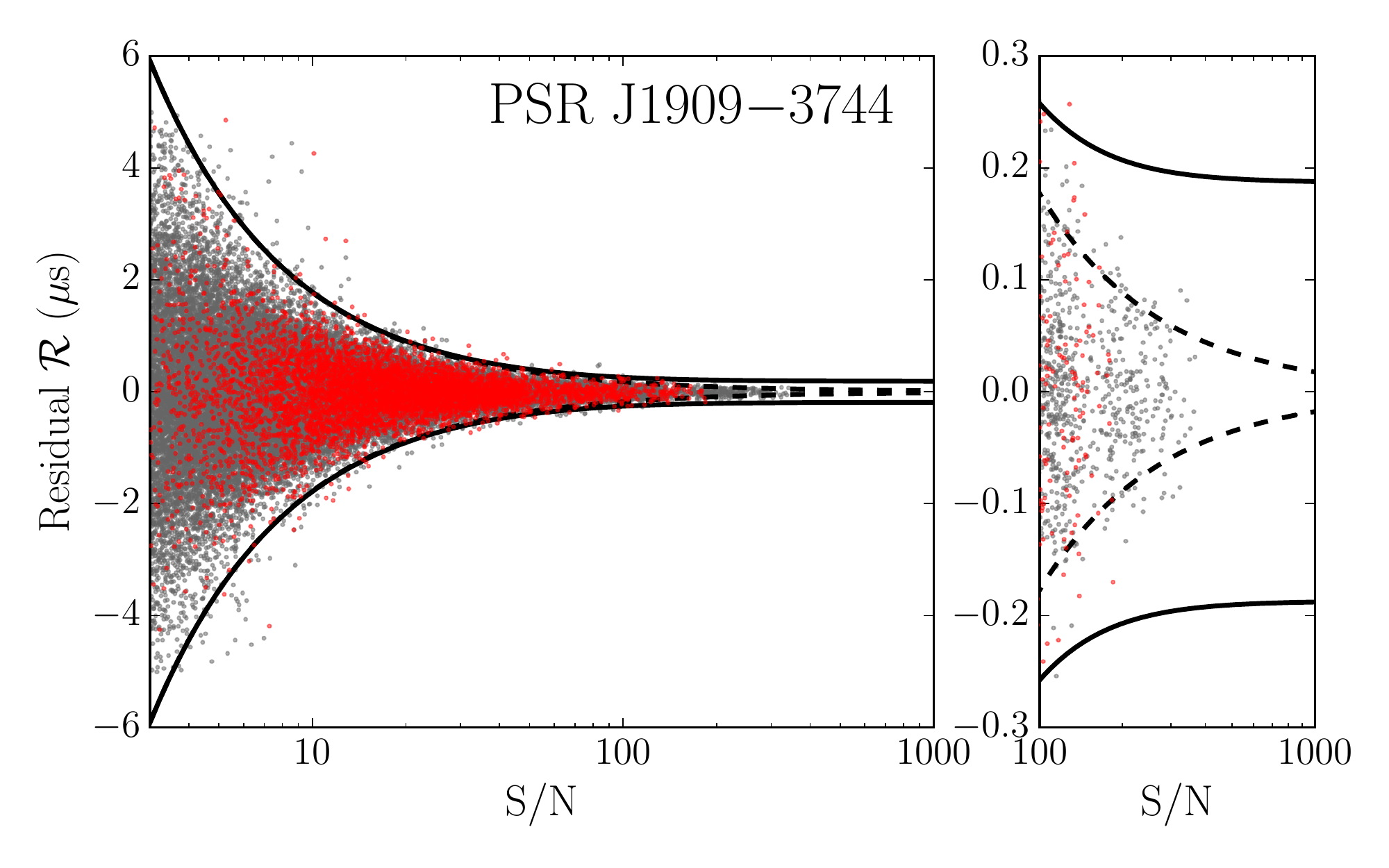}
  \caption{\footnotesize Short-term residuals $\R(\nu,t)$ vs S/N for PSR~J1909$-$3744. See Figure~\ref{fig:J1713} for more details.}
  \label{fig:J1909}
\end{figure}

\begin{figure}[ht!]
\hspace{-5ex}
\includegraphics[width=0.5\textwidth]{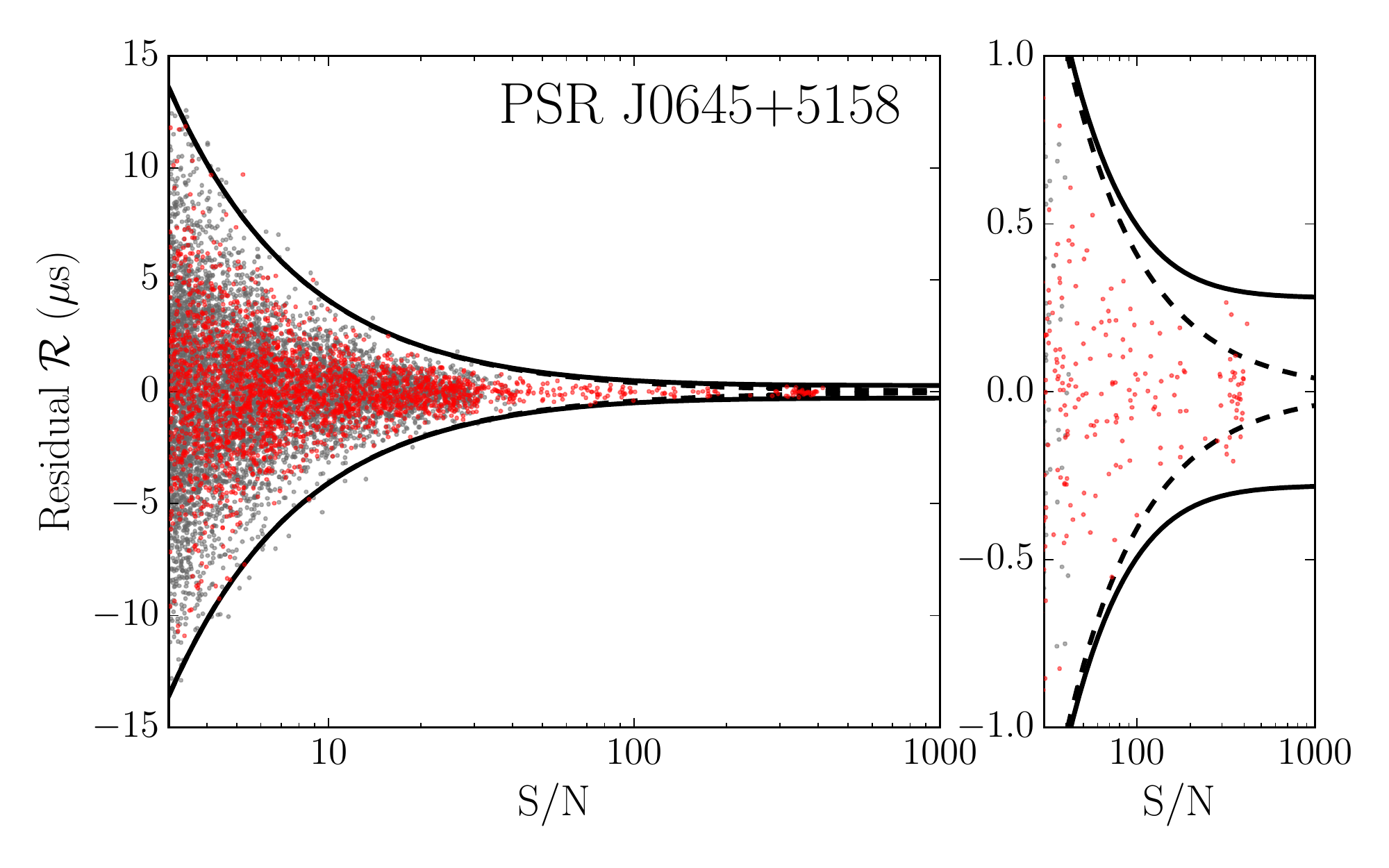}
  \caption{\footnotesize Short-term residuals $\R(\nu,t)$ vs S/N for PSR~J0645+5158. See Figure~\ref{fig:J1713} for more details.}
  \label{fig:J0645}
\end{figure}

\begin{figure}[h!]
\hspace{-5ex}
\includegraphics[width=0.5\textwidth]{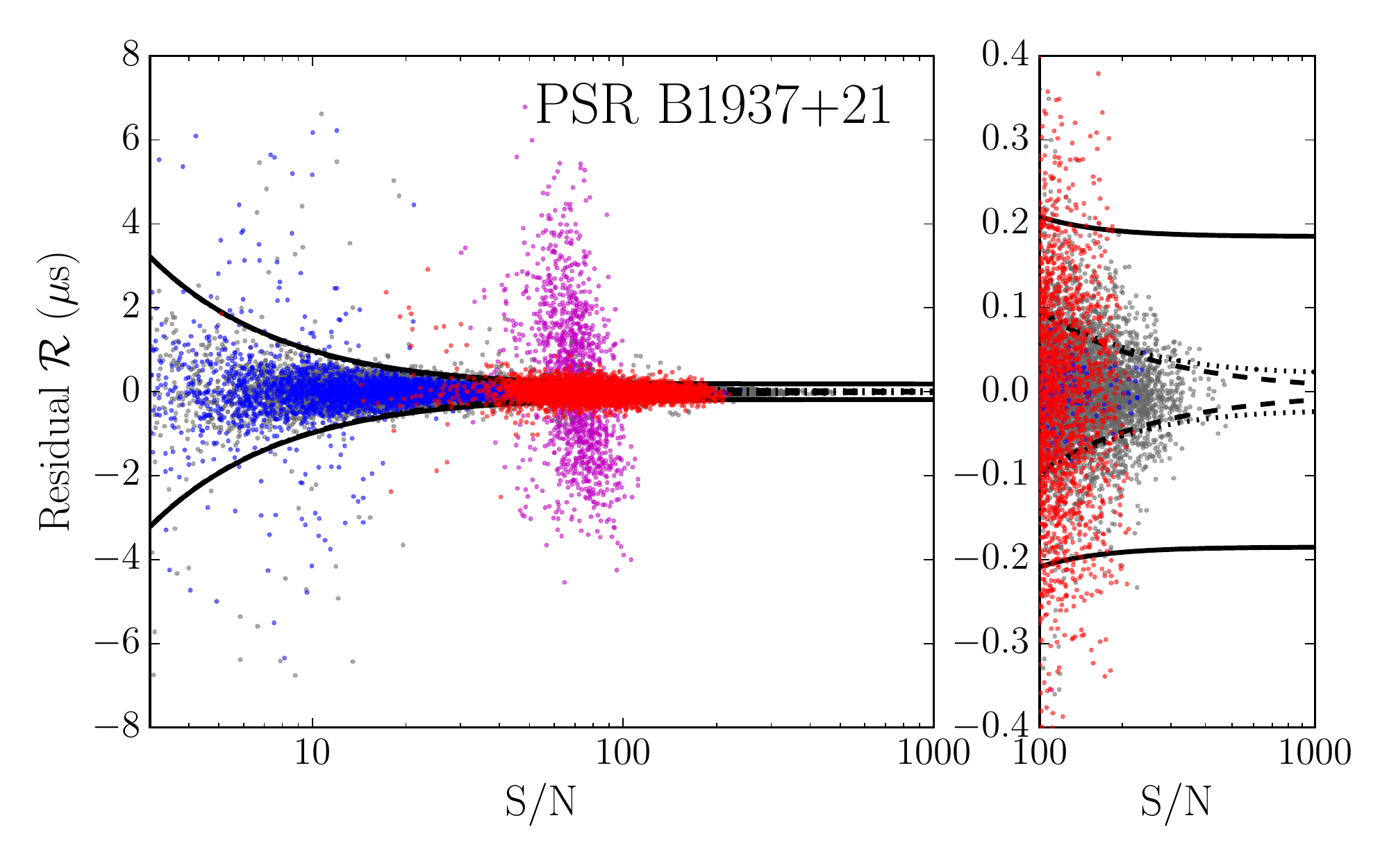}
  \caption{\footnotesize Short-term residuals $\R(\nu,t)$ vs S/N for PSR~B1937+21. See Figure~\ref{fig:J1713} for more details. The dotted line shows $\pm 3\sqrt{\sigma_{\SN}(S|\Weffref)^2 + \sigma_\DISS(1500~\mathrm{MHz}|\Dtdo,\Dnudo)^2}$, i.e., the error ranges without the inclusion of $\sigma_\J$, since it is the only pulsar of the ones we show with a visible contribution on the scale of the plot; at lower frequencies the DISS component becomes far more dominant. The magenta points show residuals in the 430~MHz band (recall that the channel bandwidth is much less and therefore the average $S$ will be lower in comparison to the other bands) and very clearly show the effect of $\sigma_{\DISS}$ increasing at lower frequencies; this effect can also be seen in the 820~MHz band residuals (red).}
  \label{fig:B1937}
\end{figure}

\begin{figure}[h!]
%\vspace{-10ex}
\hspace{-5ex}
\includegraphics[width=0.5\textwidth]{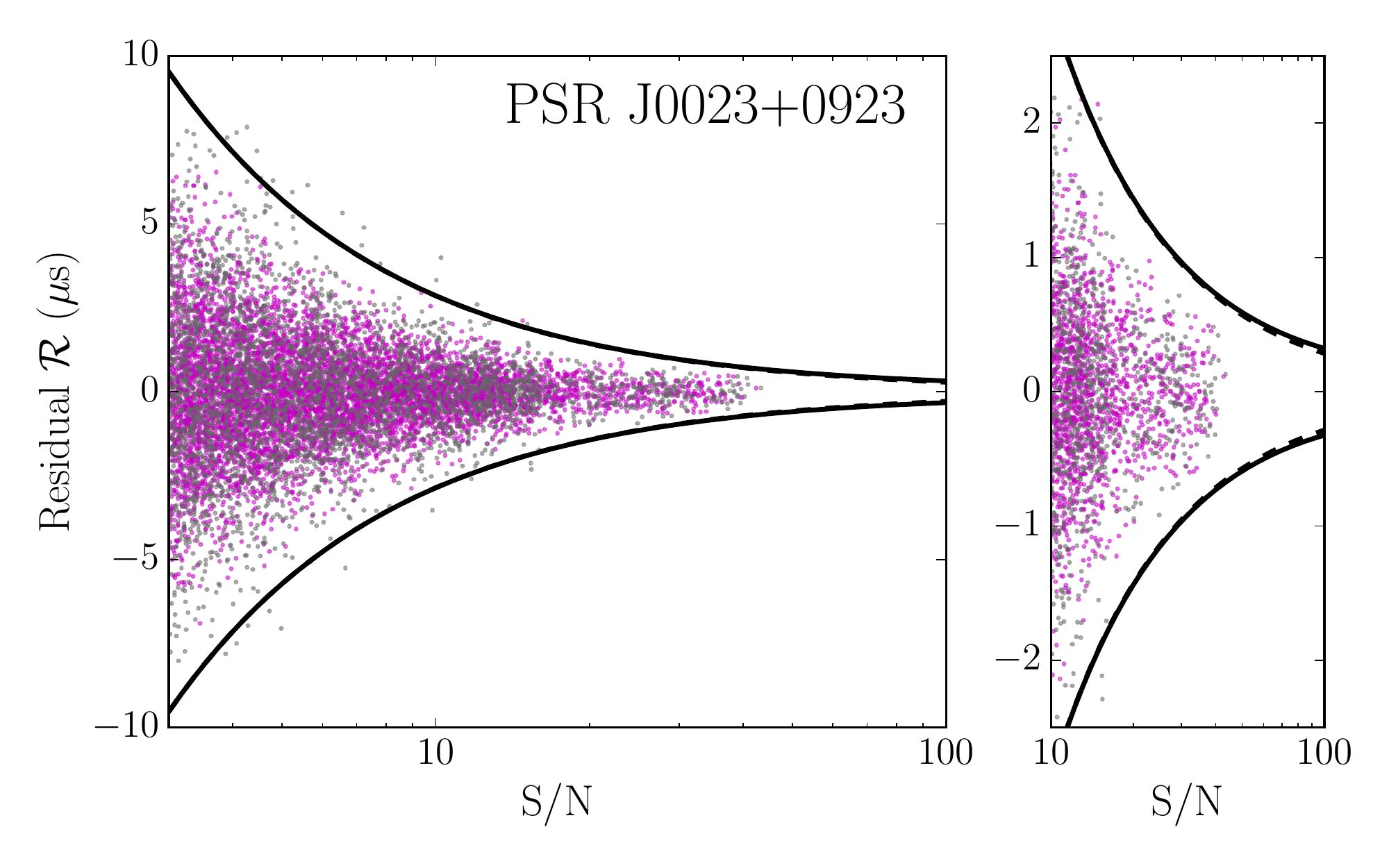}
  \caption{\footnotesize Short-term residuals $\R(\nu,t)$ vs S/N for PSR~J0023+0923. See Figure~\ref{fig:J1713} for more details. The magenta points show residuals in the 430~MHz band (recall that the channel bandwidth is much less and therefore the average $S$ will be lower in comparison to the 1500~MHz band). This pulsar shows no definitive detection of jitter. The  dashed lines show the 95\% upper limit on the constant (Model A) $\sigma_\J$, scaled to 80~s since AO was used to observe this pulsar.}
  \label{fig:J0023}
\end{figure}

\begin{deluxetable}{clc}
\tablecolumns{3}
\tablecaption{Frequency-dependent Jitter Models\label{table:models}}
\tablehead{
\colhead{\tablespace Model} & \colhead{Model} & \colhead{Single-Pulse RMS}\\
\colhead{Letter} & \colhead{Description} & \colhead{Jitter Functional Form}
}
\startdata
\tablespace A & Constant & $\sigma_{\J,1}$\\
\tablespace B & $N$-Band & $\sigma_{\J,1;\nu_0}(\nu)$ where $\nu$ is in band $\nu_0$\\
\tablespace C & Power Law & $\sigma_{\J,1,1000}(\nu/1000~\mathrm{MHz})^\alpha$ \\
\tablespace D & Power Law& $\sigma_{\J,1,1000}(\nu/1000~\mathrm{MHz})^\alpha + \sigma_{\J,1} $\\
\tablespace & plus Constant & \\
E$\kmax^{\rm a}$ & Log-Polynomial & $\sum_{k=0}^\kmax \sigma_{\J,1,k} \left[\log_{10}(\nu/1000~\mathrm{MHz})\right]^k$, $\kmax \ge 1$\\
\enddata
\tablenotetext{a}{We list these models as E1, E2, etc., depending on the number of coefficients used. }
\end{deluxetable}

The different models we considered were:
\begin{quote}
{\em A. Constant rms jitter with frequency, $\sigma_{\J,1}(\nu) = \sigma_{\J,1}$}:  
A constant-in-frequency model can leverage all of the data to obtain an estimate of the rms jitter in the case where the pulses typically have low S/N. In some cases it may be the preferred model when the frequency dependence is negligible over our observed frequency ranges.

{\em B. Per-frequency-band rms jitter, $\sigma_{\J,1;\nu_0}(\nu)$ where $\nu$ is in band $\nu_0$}: 
Similar to Model A but following the method of NG9WN, we report per-frequency-band estimates to provide somewhat finer frequency resolution to the rms jitter estimates, which allows one to compare with previous results in NG9WN and \citet{sod+2014}.

{\em C. Power-law frequency dependence, $\sigma_{\J,1}(\nu) = \sigma_{\J,1,1000}(\nu/1000~\mathrm{MHz})^\alpha$}: 
Pulse widths have been observed to roughly scale as some power of frequency over some range of frequencies \citep{Kuzmin+1986}; theoretical descriptions of the emission mechanism for canonical (slow-period) pulsars also predict such scalings \citep{rs1975,mu1989}. Width variations for MSPs are observed to follow closer to constant frequency dependence though with emission characteristics potentially similar to canonical pulsars \citep{kll+1999}. As a reminder, from the profiles used in this analysis we see that there are clear frequency dependencies for the profile component widths. As the rms jitter should be correlated to the width of the pulse (e.g., a narrow pulse shape cannot result from single pulse emission varying greatly in phase), we tested a standard power-law-dependence model. We referenced our results to 1000~MHz.

{\em D. Power-law frequency dependence plus constant, $\sigma_{\J,1}(\nu) = \sigma_{\J,1,1000}(\nu/1000~\mathrm{MHz})^\alpha + \sigma_{\J,1}$}: 
\citet{Thorsett1991} found the frequency dependence of pulse component separation in slow-period pulsars could be described in the functional form of a power law plus a constant term over a wide range of radio frequencies; the power-law index was negative for the pulsars analyzed. Similarly, \citet{Xilouris+1996} found that pulse widths could be described in the same form (this is consistent with \citet{Kuzmin+1986} for the power-law index at the highest frequencies). As with Model C, we referenced the power-law rms jitter component to 1000~MHz.

%Such a model could describe the jitter of different components between overlap versus separation in phase though could also describe the jitter dependencies of primary versus secondary components. 

% In the case where multiple pulse components are separated, the rms jitter should remain constant if the components separate further yet statistically vary between single pulses in the same way \todo{true?}.

{\em E. Log-polynomial frequency dependence, $\sigma_{\J,1}(\nu) = \sum_{k=0}^\kmax \sigma_{\J,1,k}(\log_{10}(\nu/1000~\mathrm{MHz}))^k$, $\kmax \ge 1$}:  
This model offers more flexibility and potential smoothness across frequency than the power-law dependence. For pulsars with sufficient number of residuals at high S/N, we included higher-order polynomial terms by successively increasing the value of $\kmax$ and testing for signficance using an F-test.

\end{quote}

For each pulsar, we ran our MCMC pipeline and obtained likelihood functions for each of the models. The results for each pulsar are shown in Table~\ref{table:resultsA}. We report detections of $\sigma_\J$ that are $2\sigma$ significant, listing the median and the $1\sigma$ ($\pm 34.1\%$) confidence interval. Otherwise we list the 95\% upper limits. For Models C and D we required that rms jitter values be significant at the $2\sigma$ level or we did not report the estimated parameters. Values of $\alpha$ provided in the tables are listed as the median with the $1\sigma$ confidence intervals. The last column shows the preferred model with the lowest Bayesian Information Criterion \citep[BIC;][]{BIC}, calculated as
\be
\mathrm{BIC}~=~\ln(N_{\rm resid}) N_{\rm param} - 2 \ln(\Like(\hat{\thetavec}| \left\{S_i,\R_i,\nu_i\right\},\chivec))
\label{eq:BIC}
\ee
where $N_{\rm resid}$ is the number of residuals and $N_{\rm param}$ is the number of jitter-model parameters; models with an increased number of parameters are therefore penalized. Table~\ref{table:resultsB} is in the same format as Table~\ref{table:resultsA} except that we have scaled the values of $\sigma_{\J,1}$ to $\sigma_{\J,30}$, the rms jitter for a 30-minute observation, for convenience. %Our analysis code is available at \url{https://github.com/mtlam/PulseJitterModeling}.

We show short-term residuals $\R$ for five select pulsars from the data set analyzed. Figure~\ref{fig:J1713} shows $\R$ for PSR~J1713+0747 as a function of S/N. We color the residuals by frequency band for visual clarity only and stress that our analysis keeps full frequency information for all of the residuals. The solid lines show $\pm 3\sigma_\R$ for all parameters referenced to 1500~MHz. We have taken the $\sigma_\J$ value from the constant model (Model A) and have scaled it to 120~s using Eq.~\ref{eq:sigmaJ} (note that many of the residuals were measured with AO and have 80~s subintegrations). The dashed lines show $\pm 3\sigma_\SN(S|\Weffref)$, i.e., the error ranges from the template-fitting component error only. We see that even the approximate $\sigma_\SN(S)$ does not adequately describe the increased variance of the residuals at large S/N. Figures~\ref{fig:J1909} and \ref{fig:J0645} show the residuals for PSRs~J1909$-$3744 and J0645+5158, respectively. Figure~\ref{fig:B1937} shows the residuals for PSR~B1937+21, which we observe with both telescopes (the parenthetical note about 80~s subintegrations with AO again applies). Lastly, we show the residuals for PSR~J0023+0923 in Figure~\ref{fig:J0023}, where we do not detect jitter via any model. The calculation of the lines use the 95\% upper limits on $\sigma_\J$ in the constant model; The dashed line is very barely visible against the solid line, showing our lack of detection.

We summarize the results of our single-pulsar analyses as follows.
\begin{itemize}[noitemsep,topsep=-12pt] %leftmargin=*,
\item We detect significant jitter in 43/48 pulsars.
\item We find significant frequency dependence for 30 pulsars.
\item The number of pulsars preferred for each model:\\Model A: 13,\\ Model B: 5,\\Model C: 19,\\ Model D: 1,\\ Model E2: 4,\\ Model E3: 1.
\item No pulsars show significant Model E4 parameters.
\end{itemize}

%{\bf Advantages and Disadvantages:} The disadvantage to updating the model is that it will (i) require further though hopefully not extensive code modifications, (ii) require more model parameters, (iii) pull together more prior information. However, the advantages are clear. In many cases, we found an upper limit on the jitter in one band and a detection in another; this new method will allow us to simultaneously use all of the data at hand to potentially detect jitter as well as its frequency dependence.

\subsection{Summary Statistics}
\label{sec:summary}

\begin{figure}[t!]
%\hspace{-5ex}
\vspace{-2ex}
\includegraphics[width=0.5\textwidth]{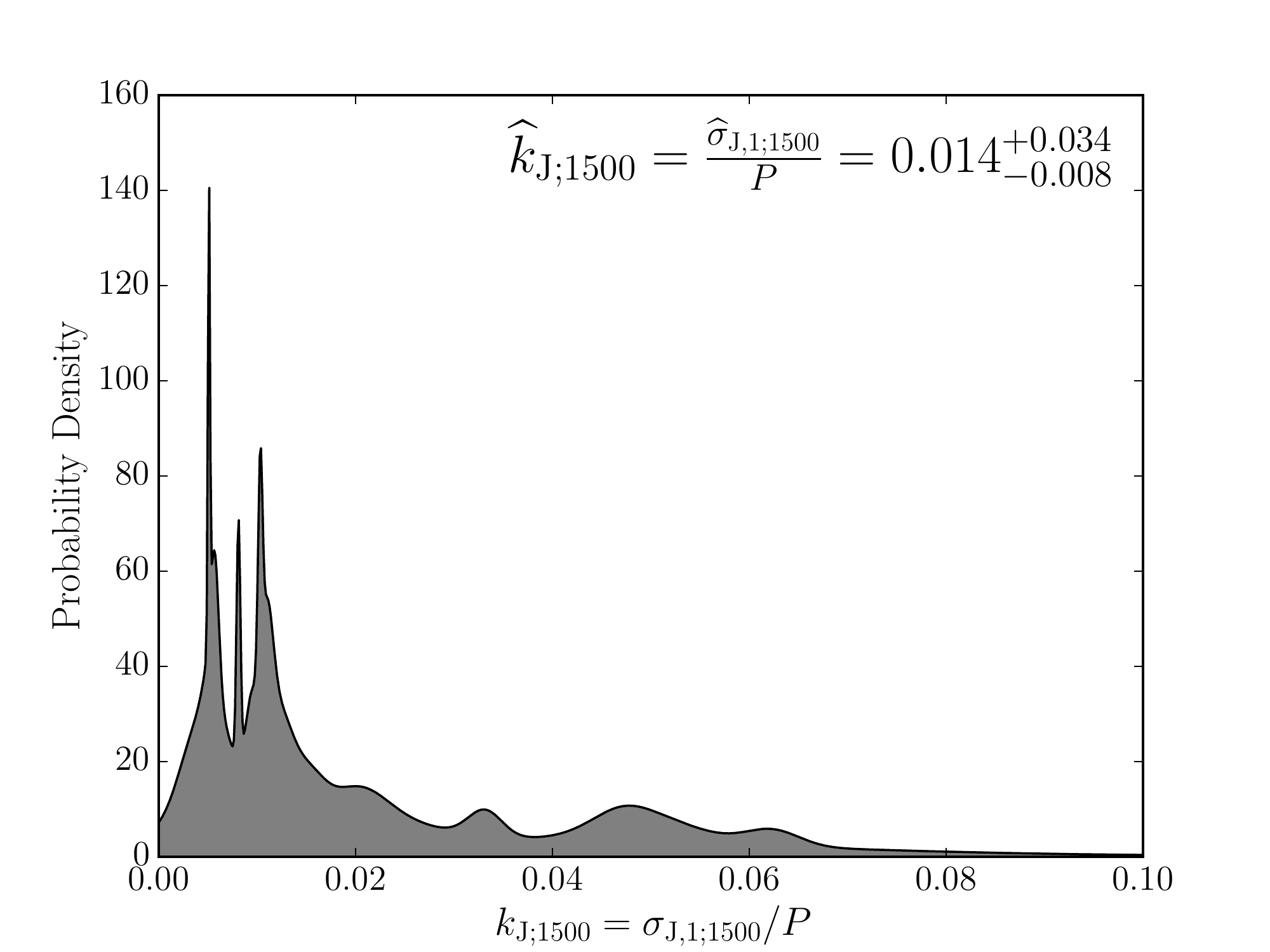}
  \caption{\footnotesize Continuous histogram of the jitter parameter $\kJref$.}
  \label{fig:kJ}
\end{figure}

Here we look at the overall statistics of jitter in our data set. Since pulse periods vary and therefore the absolute value of $\sigma_{\J,1}$ is expected to vary, we instead look at the statistics of the jitter parameter at 1500~MHz, $\kJref \equiv \sigma_{\J,1;1500}/P$. Figure~\ref{fig:kJ} shows a continuous histogram, the normalized sum of the likelihoods for $\kJref$ (see NG9WN for more details), using the preferred model per pulsar. Those values which are more constrained appear sharper with higher peaks since each likelihood has unit area. The median jitter parameter value is $\kJref = 0.014^{+0.034}_{-0.008}$, consistent with the estimates over the entire 1500~MHz band provided in NG9WN. We found that the frequency choice for $\kJ$ did not matter significantly in our analyses. For example, looking at a range of frequencies across the 1500~MHz band where all pulsars were observed, the total deviation in the estimated median $\kJ(\nu)$ was about 0.005, much lower than the scatter shown in Figure~\ref{fig:kJ}, i.e., the ensemble set for values such as $k_{\rm J;1000}$ do not vary much from $\kJref$. An analysis of individual pulsars and the frequency-dependence components may yield insights into subtle variations in the emission mechanisms between MSPs versus canonical pulsars \citep[e.g., again see][]{kll+1999}.

%\newpage

\begin{longrotatetable}
\startlongtable
\begin{deluxetable*}{l|c|cccc|cc|ccc|>{\centering\arraybackslash}p{3.75cm}|c}
\tablecolumns{14}
\tabletypesize{\scriptsize}
\tablecaption{Jitter Results Scaled to Single Pulse Rms\label{table:resultsA}}
\tablehead{
\colhead{Pulsar} & \colhead{Model A} & \multicolumn{4}{c}{Model B} & \multicolumn{2}{c}{Model C} & \multicolumn{3}{c}{Model D} & \colhead{Model E} & \colhead{Preferred}\\
\colhead{} & \colhead{(Constant)} & \multicolumn{4}{c}{($N$-Band)} & \multicolumn{2}{c}{(Power-Law)} & \multicolumn{3}{c}{(Power-Law plus Constant)} & \colhead{(Log-Polynomial)} & \colhead{Model}\\
\colhead{} & \colhead{$\sigma_{\rm J,1}$} & \colhead{$\sigma_{\rm J,1;327/430}$}  & \colhead{$\sigma_{\rm J,1;820}$}  & \colhead{$\sigma_{\rm J,1;1400}$} & \colhead{$\sigma_{\rm J,1;2300}$} & \colhead{$\sigma_{\rm J,1;1000}$} & \colhead{$\alpha$} & \colhead{$\sigma_{\rm J,1;1000}$} & \colhead{$\alpha$} & \colhead{$\sigma_{\rm J,1}$} & \colhead{$\sigma_{\rm J,1,0}, \sigma_{\rm J,1,1}, \cdots$} & \colhead{}\\
\colhead{} & \colhead{($\mu$s)} & \colhead{($\mu$s)} & \colhead{($\mu$s)} & \colhead{($\mu$s)} & \colhead{($\mu$s)} & \colhead{($\mu$s)} & \colhead{} & \colhead{($\mu$s)} & \colhead{} & \colhead{($\mu$s)} & \colhead{($\mu$s)} & \colhead{}
}
\startdata
\vspace{-1ex} J0023+0923 & $<$8.2 & $<$7.7 & $\cdots$ & $20_{-10}^{+8}$ & $\cdots$ & 2.5$_{-1.1}^{+2.0}$ & $7.8_{-1.3}^{+1.2}$ & $\cdots$ & $\cdots$ & $\cdots$ & $\cdots$ & C (PL)\\
\vspace{-1ex} J0030+0451 & 131.7$_{-2.9}^{+3.0}$ & $\cdots$ & $\cdots$ & 52.6$ \pm 3.0$ & 207.1$ \pm 3.2$ & 141.3$_{-2.8}^{+2.5}$ & 1.07$ \pm 0.04$ & $\cdots$ & $\cdots$ & $\cdots$ & $\cdots$ & B ($N$-Band)\\
\vspace{-1ex} J0340+4130 & $<$270 & $\cdots$ & $181_{-83}^{+62}$ & $<$460 & $\cdots$ & $\cdots$ & $\cdots$ & $\cdots$ & $\cdots$ & $\cdots$ & $\cdots$ & $\cdots$\\
\vspace{-1ex} J0613$-$0200 & 47.0$_{-2.6}^{+2.4}$ & $\cdots$ & 43.7$_{-3.1}^{+2.7}$ & $143_{-9}^{+8}$ & $\cdots$ & 66.5$ \pm 2.9$ & $1.7 \pm 0.1$ & $\cdots$ & $\cdots$ & $\cdots$ & 77.7$_{-5.5}^{+4.1}$,$360_{-60}^{+40}$ & B ($N$-Band)\\
\vspace{-1ex} J0636+5128 & $58_{-11}^{+10}$ & $\cdots$ & $52_{-12}^{+14}$ & $69_{-24}^{+17}$ & $\cdots$ & $52_{-15}^{+12}$ & $0.2_{-0.9}^{+0.7}$ & $\cdots$ & $\cdots$ & $\cdots$ & $\cdots$ & A (Const)\\
\vspace{-1ex} J0645+5158 & 10.8$_{-1.3}^{+1.6}$ & $\cdots$ & 10.6$_{-1.5}^{+1.3}$ & 22.3$_{-6.0}^{+4.9}$ & $\cdots$ & 13.9$_{-1.4}^{+1.5}$ & $2.7_{-0.4}^{+0.3}$ & 9.9$_{-1.2}^{+1.3}$ & $9.0_{-1.1}^{+0.7}$ & 0.6$_{-0.2}^{+0.5}$ & 16.6$_{-2.5}^{+1.9}$,$69_{-27}^{+18}$ & D (PLC)\\
\vspace{-1ex} J0740+6620 & $<$31 & $\cdots$ & $<$31 & $<$51 & $\cdots$ & $\cdots$ & $\cdots$ & $\cdots$ & $\cdots$ & $\cdots$ & $\cdots$ & $\cdots$\\
\vspace{-1ex} J0931$-$1902 & $<$94 & $\cdots$ & $<$130 & $<$97 & $\cdots$ & $\cdots$ & $\cdots$ & $\cdots$ & $\cdots$ & $\cdots$ & $\cdots$ & $\cdots$\\
\vspace{-1ex} J1012+5307 & 26.0$_{-4.7}^{+4.2}$ & $\cdots$ & $<$17.8 & 55.4$_{-4.8}^{+4.5}$ & $\cdots$ & 22.8$_{-4.0}^{+3.9}$ & $2.3 \pm 0.4$ & $\cdots$ & $\cdots$ & $\cdots$ & 15.7$_{-4.5}^{+6.2}$,$271_{-45}^{+33}$ & E2 (LP2)\\
\vspace{-1ex} J1022+1001 & 126.3$ \pm 1.7$ & $\cdots$ & $\cdots$ & 40.5$_{-5.8}^{+4.7}$ & 145.4$_{-2.3}^{+2.2}$ & 113.5$_{-2.3}^{+2.2}$ & 0.36$_{-0.03}^{+0.04}$ & $\cdots$ & $\cdots$ & $\cdots$ & 111.2$_{-2.8}^{+2.1}$,$125_{-12}^{+9}$ & E2 (LP2)\\
\vspace{-1ex} J1024$-$0719 & $<$34 & $\cdots$ & $<$20 & $111 \pm 10$ & $\cdots$ & 14.8$_{-4.8}^{+5.9}$ & $4.9_{-0.7}^{+0.8}$ & $\cdots$ & $\cdots$ & $\cdots$ & $\cdots$ & C (PL)\\
\vspace{-1ex} J1125+7819 & $124_{-15}^{+14}$ & $\cdots$ & $58_{-28}^{+23}$ & $212 \pm 18$ & $\cdots$ & $\cdots$ & $\cdots$ & $\cdots$ & $\cdots$ & $\cdots$ & $\cdots$ & B ($N$-Band)\\
\vspace{-1ex} J1453+1902 & $265 \pm 28$ & $327_{-90}^{+88}$ & $\cdots$ & $257_{-31}^{+30}$ & $\cdots$ & $\cdots$ & $\cdots$ & $\cdots$ & $\cdots$ & $\cdots$ & $\cdots$ & A (Const)\\
\vspace{-1ex} J1455$-$3330 & $101 \pm 12$ & $\cdots$ & $<$64 & $156_{-13}^{+12}$ & $\cdots$ & $95_{-12}^{+11}$ & $0.5_{-0.3}^{+0.2}$ & $\cdots$ & $\cdots$ & $\cdots$ & $\cdots$ & A (Const)\\
\vspace{-1ex} J1600$-$3053 & $<$26 & $\cdots$ & $<$20 & $24_{-7}^{+6}$ & $\cdots$ & 0.3$_{-0.1}^{+0.6}$ & $9.0_{-2.0}^{+0.8}$ & $\cdots$ & $\cdots$ & $\cdots$ & $\cdots$ & C (PL)\\
\vspace{-1ex} J1614$-$2230 & $64_{-8}^{+7}$ & $\cdots$ & $<$56 & $74 \pm 8$ & $\cdots$ & $37_{-16}^{+15}$ & $1.5_{-0.8}^{+1.2}$ & $\cdots$ & $\cdots$ & $\cdots$ & $\cdots$ & A (Const)\\
\vspace{-1ex} J1640+2224 & 27.9$ \pm 1.4$ & $<$4.1 & $\cdots$ & 51.3$_{-2.1}^{+1.9}$ & $\cdots$ & 8.5$ \pm 1.4$ & $5.1_{-0.3}^{+0.4}$ & $\cdots$ & $\cdots$ & $\cdots$ & 38.1$_{-1.6}^{+1.7}$,$111_{-7}^{+6}$ & C (PL)\\
\vspace{-1ex} J1643$-$1224 & $<$26 & $\cdots$ & $<$16.5 & $105_{-8}^{+7}$ & $\cdots$ & $27_{-10}^{+9}$ & $2.0_{-0.5}^{+0.7}$ & $\cdots$ & $\cdots$ & $\cdots$ & $\cdots$ & C (PL)\\
\vspace{-1ex} J1713+0747 & 32.0$ \pm 0.1$ & $\cdots$ & 67.3$_{-22.9}^{+1.5}$ & 26.0$_{-0.8}^{+2.8}$ & 25.3$_{-5.2}^{+0.7}$ & 43.6$_{-0.4}^{+0.3}$ & $-$1.16$ \pm 0.02$ & 22.5$ \pm 0.2$ & $-5.0 \pm 0.1$ & 18.5$ \pm 0.6$ & 40.4$ \pm 0.3$,$-$65.7$_{-1.0}^{+1.3}$ & B ($N$-Band)\\
\vspace{-1ex} J1738+0333 & 31.4$_{-4.9}^{+4.3}$ & $\cdots$ & $\cdots$ & 36.8$_{-4.7}^{+4.4}$ & $<$22 & $54_{-12}^{+18}$ & $-1.7_{-1.2}^{+0.8}$ & $\cdots$ & $\cdots$ & $\cdots$ & $\cdots$ & A (Const)\\
\vspace{-1ex} J1741+1351 & 41.5$_{-1.9}^{+2.0}$ & $<$20 & $\cdots$ & 43.5$ \pm 2.2$ & $\cdots$ & 30.3$_{-3.8}^{+3.7}$ & $1.0_{-0.3}^{+0.4}$ & $\cdots$ & $\cdots$ & $\cdots$ & 31.9$_{-3.0}^{+2.9}$,$75_{-17}^{+18}$ & E2 (LP2)\\
\vspace{-1ex} J1744$-$1134 & 42.0$ \pm 0.9$ & $\cdots$ & 39.3$_{-1.6}^{+1.7}$ & 43.7$ \pm 1.2$ & $\cdots$ & 41.2$_{-1.2}^{+1.0}$ & 0.12$ \pm 0.08$ & $\cdots$ & $\cdots$ & $\cdots$ & $\cdots$ & A (Const)\\
\vspace{-1ex} J1832$-$0836 & $<$57 & $\cdots$ & $<$940 & $<$60 & $\cdots$ & $\cdots$ & $\cdots$ & $\cdots$ & $\cdots$ & $\cdots$ & $\cdots$ & $\cdots$\\
\vspace{-1ex} J1853+1303 & $62 \pm 6$ & $<$240 & $\cdots$ & $63_{-7}^{+6}$ & $\cdots$ & $74_{-14}^{+11}$ & $-0.5_{-0.4}^{+0.5}$ & $\cdots$ & $\cdots$ & $\cdots$ & $\cdots$ & A (Const)\\
\vspace{-1ex} B1855+09 & 115.8$ \pm 1.3$ & $\cdots$ & $\cdots$ & $<$240 & 116.3$ \pm 1.7$ & 105.6$_{-5.2}^{+3.1}$ & 0.25$_{-0.07}^{+0.13}$ & $\cdots$ & $\cdots$ & $\cdots$ & 107.1$ \pm 2.8$,$54_{-16}^{+15}$ & C (PL)\\
\vspace{-1ex} J1903+0327 & $<$110 & $\cdots$ & $\cdots$ & $<$110 & $<$170 & $837_{-228}^{+124}$ & $-5.1_{-0.8}^{+0.7}$ & $\cdots$ & $\cdots$ & $\cdots$ & $\cdots$ & C (PL)\\
\vspace{-1ex} J1909$-$3744 & 12.6$ \pm 0.3$ & $\cdots$ & 17.8$_{-0.8}^{+0.9}$ & 11.7$ \pm 0.3$ & $\cdots$ & 16.0$_{-0.5}^{+0.4}$ & $-$0.87$ \pm 0.09$ & $\cdots$ & $\cdots$ & $\cdots$ & 16.3$ \pm 0.5$,$-$28.7$_{-3.3}^{+2.9}$ & C (PL)\\
\vspace{-1ex} J1910+1256 & 84.3$_{-4.0}^{+3.8}$ & $\cdots$ & $\cdots$ & 82.6$_{-4.1}^{+4.2}$ & $102_{-16}^{+15}$ & $45_{-6}^{+7}$ & $1.3 \pm 0.3$ & $\cdots$ & $\cdots$ & $\cdots$ & $\cdots$ & C (PL)\\
\vspace{-1ex} J1911+1347 & 38.3$_{-2.1}^{+2.0}$ & $\cdots$ & $\cdots$ & 38.4$_{-2.0}^{+2.1}$ & $<$62 & $93_{-11}^{+14}$ & $-2.5_{-0.5}^{+0.4}$ & $\cdots$ & $\cdots$ & $\cdots$ & $74 \pm 7$,$-222 \pm 44$ & C (PL)\\
\vspace{-1ex} J1918$-$0642 & $36_{-7}^{+6}$ & $\cdots$ & $<$44 & $40_{-8}^{+6}$ & $\cdots$ & 0.6$_{-0.3}^{+3.5}$ & $8.5_{-3.4}^{+1.1}$ & $\cdots$ & $\cdots$ & $\cdots$ & $\cdots$ & C (PL)\\
\vspace{-1ex} J1923+2515 & $196 \pm 8$ & $<$210 & $\cdots$ & $199_{-8}^{+7}$ & $\cdots$ & $213 \pm 9$ & $-$0.36$ \pm 0.09$ & $\cdots$ & $\cdots$ & $\cdots$ & $\cdots$ & C (PL)\\
\vspace{-1ex} B1937+21 & 17.0$ \pm 0.2$ & $<$47 & 31.9$_{-5.7}^{+0.8}$ & 15.5$ \pm 0.2$ & 16.4$_{-7.5}^{+1.4}$ & 23.4$ \pm 0.3$ & $-$1.02$_{-0.04}^{+0.03}$ & 12.3$_{-0.5}^{+0.4}$ & $-3.0 \pm 0.2$ & 9.8$_{-0.6}^{+0.8}$ & 22.5$ \pm 0.3$,$-$36.4$ \pm 1.6$ \newline 22.7$ \pm 0.3$,$-$84.2$_{-3.5}^{+4.8}$,$219_{-18}^{+13}$ & E3 (LP3)\\
\vspace{-1ex} J1944+0907 & $251 \pm 5$ & $<$100 & $\cdots$ & $259 \pm 5$ & $\cdots$ & $121_{-10}^{+9}$ & $2.1 \pm 0.2$ & $\cdots$ & $\cdots$ & $\cdots$ & $\cdots$ & C (PL)\\
\vspace{-1ex} J1946+3417$^{\rm a}$  & $197 \pm 8$ & $\cdots$ & $\cdots$ & $197 \pm 9$ & $\cdots$ & $330 \pm 50$ & $-1.3 \pm 0.4$ & $\cdots$ & $\cdots$ & $\cdots$ & $\cdots$ & A (Const)\\
\vspace{-1ex} B1953+29 & $330_{-9}^{+8}$ & $436_{-20}^{+22}$ & $\cdots$ & $309 \pm 10$ & $\cdots$ & $343_{-9}^{+8}$ & $-$0.32$ \pm 0.05$ & $\cdots$ & $\cdots$ & $\cdots$ & $\cdots$ & C (PL)\\
\vspace{-1ex} J2010$-$1323 & $60_{-11}^{+8}$ & $\cdots$ & $<$43 & $104 \pm 11$ & $\cdots$ & $47_{-13}^{+11}$ & $1.8_{-0.5}^{+0.6}$ & $\cdots$ & $\cdots$ & $\cdots$ & $\cdots$ & C (PL)\\
\vspace{-1ex} J2017+0603 & $24 \pm 5$ & $\cdots$ & $\cdots$ & $<$510 & $21_{-8}^{+5}$ & 6.6$_{-3.0}^{+3.2}$ & $3.3_{-0.6}^{+0.7}$ & $\cdots$ & $\cdots$ & $\cdots$ & $\cdots$ & C (PL)\\
\vspace{-1ex} J2033+1734 & $380 \pm 12$ & $331_{-19}^{+21}$ & $\cdots$ & $410_{-16}^{+17}$ & $\cdots$ & $390 \pm 13$ & 0.25$_{-0.06}^{+0.07}$ & $\cdots$ & $\cdots$ & $\cdots$ & $\cdots$ & C (PL)\\
\vspace{-1ex}J2043+1711 & 15.7$_{-3.8}^{+3.5}$ & $<$12.6 & $\cdots$ & 22.2$ \pm 3.2$ & $\cdots$ & $\cdots$ & $\cdots$ & $\cdots$ & $\cdots$ & $\cdots$ & $\cdots$ & A (Const)\vspace{2ex}\\
\vspace{-1ex} J2145$-$0750 & 113.3$ \pm 1.0$ & $\cdots$ & 130.5$_{-1.5}^{+1.7}$ & 81.9$_{-1.7}^{+1.5}$ & $\cdots$ & 110.3$_{-1.7}^{+1.3}$ & $-$0.76$ \pm 0.05$ & $62_{-8}^{+6}$ & $-2.1_{-0.5}^{+0.4}$ & $43_{-7}^{+8}$ & $\cdots$ & B ($N$-Band)\\
\vspace{-1ex} J2214+3000 & $103 \pm 5$ & $\cdots$ & $\cdots$ & 103.3$_{-5.1}^{+5.0}$ & $<$150 & $83_{-13}^{+14}$ & $0.5 \pm 0.4$ & $\cdots$ & $\cdots$ & $\cdots$ & $\cdots$ & A (Const)\\
\vspace{-1ex} J2229+2643 & $113 \pm 7$ & $<$57 & $\cdots$ & $127_{-8}^{+9}$ & $\cdots$ & $89_{-11}^{+12}$ & $0.8 \pm 0.3$ & $\cdots$ & $\cdots$ & $\cdots$ & $\cdots$ & C (PL)\\
\vspace{-1ex} J2234+0611 & 19.6$_{-1.6}^{+1.7}$ & $\cdots$ & $\cdots$ & $<$98 & 19.8$_{-1.6}^{+1.5}$ & 29.8$_{-3.2}^{+3.1}$ & $-1.2 \pm 0.3$ & $\cdots$ & $\cdots$ & $\cdots$ & $40 \pm 5$,$-145_{-42}^{+38}$ & E2 (LP2)\\
\vspace{-1ex} J2234+0944 & 40.1$ \pm 2.4$ & $\cdots$ & $\cdots$ & $427_{-149}^{+258}$ & 40.3$_{-2.4}^{+2.1}$ & $68_{-13}^{+10}$ & $-1.4_{-0.4}^{+0.6}$ & $\cdots$ & $\cdots$ & $\cdots$ & $\cdots$ & A (Const)\\
\vspace{-1ex} J2302+4442 & $270_{-25}^{+24}$ & $\cdots$ & $238 \pm 36$ & $300_{-32}^{+31}$ & $\cdots$ & $255_{-29}^{+28}$ & $0.2 \pm 0.3$ & $\cdots$ & $\cdots$ & $\cdots$ & $\cdots$ & A (Const)\\
\vspace{-1ex} J2317+1439$^{\rm b}$  & 25.2$_{-2.9}^{+2.8}$ & {$<$20/$<$8.3} & $\cdots$ & 65.8$_{-3.0}^{+3.5}$ & $\cdots$ & 5.1$_{-2.0}^{+2.9}$ & $6.0 \pm 0.9$ & $\cdots$ & $\cdots$ & $\cdots$ & 46.4$_{-2.5}^{+2.7}$,$124 \pm 8$ & C (PL)\\
\vspace{-1ex} J2322+2057 & $26_{-12}^{+9}$ & $<$41 & $\cdots$ & $33_{-10}^{+9}$ & $\cdots$ & $\cdots$ & $\cdots$ & $\cdots$ & $\cdots$ & $\cdots$ & $\cdots$ & A (Const)\vspace{1ex}\\
\enddata
\tablenotetext{a}{\vspace{-2ex}No residuals with $S > 3$.}
\tablenotetext{b}{We list the rms jitter for the 327~MHz band followed by the 430~MHz band in the same column for Model B.}
\end{deluxetable*}
\end{longrotatetable}

%\newpage

\begin{longrotatetable}
\startlongtable
\begin{deluxetable*}{l|c|cccc|cc|ccc|>{\centering\arraybackslash}p{3.75cm}|c}
\tablecolumns{14}
\tabletypesize{\scriptsize}
\tablecaption{Jitter Results Scaled to 30 Minutes\label{table:resultsB}}
\tablehead{
\colhead{Pulsar} & \colhead{Model A} & \multicolumn{4}{c}{Model B} & \multicolumn{2}{c}{Model C} & \multicolumn{3}{c}{Model D} & \colhead{Model E} & \colhead{Preferred}\\
\colhead{} & \colhead{(Constant)} & \multicolumn{4}{c}{($N$-Band)} & \multicolumn{2}{c}{(Power-Law)} & \multicolumn{3}{c}{(Power-Law plus Constant)} & \colhead{(Log-Polynomial)} & \colhead{Model}\\
\colhead{} & \colhead{$\sigma_{\rm J,30}$} & \colhead{$\sigma_{\rm J,30;327/430}$}  & \colhead{$\sigma_{\rm J,30;820}$}  & \colhead{$\sigma_{\rm J,30;1400}$} & \colhead{$\sigma_{\rm J,30;2300}$} & \colhead{$\sigma_{\rm J,30;1000}$} & \colhead{$\alpha$} & \colhead{$\sigma_{\rm J,30;1000}$} & \colhead{$\alpha$} & \colhead{$\sigma_{\rm J,30}$} & \colhead{$\sigma_{\rm J,30,0}, \sigma_{\rm J,30,1}, \cdots$} & \colhead{}\\
\colhead{} & \colhead{(ns)} & \colhead{(ns)} & \colhead{(ns)} & \colhead{(ns)} & \colhead{(ns)} & \colhead{(ns)} & \colhead{} & \colhead{(ns)} & \colhead{} & \colhead{(ns)} & \colhead{(ns)} & \colhead{}
}
\startdata
\vspace{-1ex} J0023+0923 & $<$11 & $<$10 & $\cdots$ & 27$_{-13}^{+12}$ & $\cdots$ & 3.3$_{-1.5}^{+2.7}$ & $7.8_{-1.3}^{+1.2}$ & $\cdots$ & $\cdots$ & $\cdots$ & $\cdots$ & C (PL)\\
\vspace{-1ex} J0030+0451 & 216.5$_{-4.8}^{+4.9}$ & $\cdots$ & $\cdots$ & 86.5$_{-4.9}^{+5.0}$ & 340.6$ \pm 5.3$ & 232.3$_{-4.6}^{+4.2}$ & 1.07$ \pm 0.04$ & $\cdots$ & $\cdots$ & $\cdots$ & $\cdots$ & B ($N$-Band)\\
\vspace{-1ex} J0340+4130 & $<$360 & $\cdots$ & 250$_{-110}^{+90}$ & $<$630 & $\cdots$ & $\cdots$ & $\cdots$ & $\cdots$ & $\cdots$ & $\cdots$ & $\cdots$ & $\cdots$\\
\vspace{-1ex} J0613$-$0200 & 61.3$_{-3.4}^{+3.1}$ & $\cdots$ & 56.9$_{-4.0}^{+3.5}$ & 188$_{-12}^{+11}$ & $\cdots$ & 86.7$_{-3.7}^{+3.8}$ & $1.7 \pm 0.1$ & $\cdots$ & $\cdots$ & $\cdots$ & 77.7$_{-5.5}^{+4.1}$,$360_{-60}^{+40}$ & B ($N$-Band)\\
\vspace{-1ex} J0636+5128 & 74$_{-14}^{+13}$ & $\cdots$ & 66$_{-16}^{+19}$ & 87$_{-31}^{+22}$ & $\cdots$ & 67$_{-19}^{+15}$ & $0.2_{-0.9}^{+0.7}$ & $\cdots$ & $\cdots$ & $\cdots$ & $\cdots$ & A (Const)\\
\vspace{-1ex} J0645+5158 & 24.0$_{-3.0}^{+3.5}$ & $\cdots$ & 23.5$_{-3.2}^{+2.9}$ & 50$_{-13}^{+11}$ & $\cdots$ & 30.9$_{-3.1}^{+3.3}$ & $2.7_{-0.4}^{+0.3}$ & 21.9$_{-2.7}^{+3.0}$ & $9.0_{-1.1}^{+0.7}$ & 1.4$_{-0.5}^{+1.1}$ & 16.6$_{-2.5}^{+1.9}$,$69_{-27}^{+18}$ & D (PLC)\\
\vspace{-1ex} J0740+6620 & $<$39 & $\cdots$ & $<$39 & $<$65 & $\cdots$ & $\cdots$ & $\cdots$ & $\cdots$ & $\cdots$ & $\cdots$ & $\cdots$ & $\cdots$\\
\vspace{-1ex} J0931$-$1902 & $<$150 & $\cdots$ & $<$210 & $<$160 & $\cdots$ & $\cdots$ & $\cdots$ & $\cdots$ & $\cdots$ & $\cdots$ & $\cdots$ & $\cdots$\\
\vspace{-1ex} J1012+5307 & 45$_{-8}^{+7}$ & $\cdots$ & $<$30 & 95$ \pm 8$ & $\cdots$ & 39$ \pm 7$ & $2.3 \pm 0.4$ & $\cdots$ & $\cdots$ & $\cdots$ & 15.7$_{-4.5}^{+6.2}$,$271_{-45}^{+33}$ & E2 (LP2)\\
\vspace{-1ex} J1022+1001 & 381.9$ \pm 5.2$ & $\cdots$ & $\cdots$ & 122$_{-18}^{+14}$ & 440$ \pm 7$ & 343$ \pm 7$ & 0.36$_{-0.03}^{+0.04}$ & $\cdots$ & $\cdots$ & $\cdots$ & 111.2$_{-2.8}^{+2.1}$,$125_{-12}^{+9}$ & E2 (LP2)\\
\vspace{-1ex} J1024$-$0719 & $<$57 & $\cdots$ & $<$34 & 189$ \pm 18$ & $\cdots$ & 25$_{-8}^{+10}$ & $4.9_{-0.7}^{+0.8}$ & $\cdots$ & $\cdots$ & $\cdots$ & $\cdots$ & C (PL)\\
\vspace{-1ex} J1125+7819 & 190$_{-24}^{+22}$ & $\cdots$ & 89$_{-44}^{+36}$ & 324$_{-28}^{+29}$ & $\cdots$ & $\cdots$ & $\cdots$ & $\cdots$ & $\cdots$ & $\cdots$ & $\cdots$ & B ($N$-Band)\\
\vspace{-1ex} J1453+1902 & 480$ \pm 50$ & 590$ \pm 160$ & $\cdots$ & 460$_{-60}^{+50}$ & $\cdots$ & $\cdots$ & $\cdots$ & $\cdots$ & $\cdots$ & $\cdots$ & $\cdots$ & A (Const)\\
\vspace{-1ex} J1455$-$3330 & 213$_{-27}^{+26}$ & $\cdots$ & $<$130 & 330$_{-28}^{+27}$ & $\cdots$ & 202$_{-25}^{+24}$ & $0.5_{-0.3}^{+0.2}$ & $\cdots$ & $\cdots$ & $\cdots$ & $\cdots$ & A (Const)\\
\vspace{-1ex} J1600$-$3053 & $<$37 & $\cdots$ & $<$28 & 35$_{-11}^{+10}$ & $\cdots$ & 0.4$_{-0.2}^{+0.9}$ & $9.0_{-2.0}^{+0.8}$ & $\cdots$ & $\cdots$ & $\cdots$ & $\cdots$ & C (PL)\\
\vspace{-1ex} J1614$-$2230 & 86$_{-11}^{+10}$ & $\cdots$ & $<$74 & 98$_{-12}^{+11}$ & $\cdots$ & 49$_{-22}^{+20}$ & $1.5_{-0.8}^{+1.2}$ & $\cdots$ & $\cdots$ & $\cdots$ & $\cdots$ & A (Const)\\
\vspace{-1ex} J1640+2224 & 37.0$ \pm 1.9$ & $<$5 & $\cdots$ & 68.0$_{-2.8}^{+2.6}$ & $\cdots$ & 11.3$ \pm 1.8$ & $5.1_{-0.3}^{+0.4}$ & $\cdots$ & $\cdots$ & $\cdots$ & 38.1$_{-1.6}^{+1.7}$,$111_{-7}^{+6}$ & C (PL)\\
\vspace{-1ex} J1643$-$1224 & $<$41 & $\cdots$ & $<$26 & 169$_{-14}^{+12}$ & $\cdots$ & 44$_{-17}^{+15}$ & $2.0_{-0.5}^{+0.7}$ & $\cdots$ & $\cdots$ & $\cdots$ & $\cdots$ & C (PL)\\
\vspace{-1ex} J1713+0747 & 51.0$ \pm 0.2$ & $\cdots$ & 107.2$_{-36.5}^{+2.4}$ & 41.4$_{-1.3}^{+4.5}$ & 40.3$_{-8.2}^{+1.2}$ & 69.4$_{-0.6}^{+0.5}$ & $-$1.16$ \pm 0.02$ & 35.9$ \pm 0.4$ & $-5.0 \pm 0.1$ & 29.5$ \pm 0.9$ & 40.4$ \pm 0.3$,$-$65.7$_{-1.0}^{+1.3}$ & B ($N$-Band)\\
\vspace{-1ex} J1738+0333 & 57$_{-9}^{+8}$ & $\cdots$ & $\cdots$ & 66$_{-9}^{+8}$ & $<$39 & 98$_{-23}^{+33}$ & $-1.7_{-1.2}^{+0.8}$ & $\cdots$ & $\cdots$ & $\cdots$ & $\cdots$ & A (Const)\\
\vspace{-1ex} J1741+1351 & 59.9$_{-2.7}^{+2.9}$ & $<$28 & $\cdots$ & 62.7$_{-3.2}^{+3.1}$ & $\cdots$ & 43.8$_{-5.5}^{+5.4}$ & $1.0_{-0.3}^{+0.4}$ & $\cdots$ & $\cdots$ & $\cdots$ & 31.9$_{-3.0}^{+2.9}$,$75_{-17}^{+18}$ & E2 (LP2)\\
\vspace{-1ex} J1744$-$1134 & 63.2$_{-1.4}^{+1.3}$ & $\cdots$ & 59.2$_{-2.4}^{+2.6}$ & 65.7$ \pm 1.8$ & $\cdots$ & 62.1$_{-1.7}^{+1.6}$ & 0.12$ \pm 0.08$ & $\cdots$ & $\cdots$ & $\cdots$ & $\cdots$ & A (Const)\\
\vspace{-1ex} J1832$-$0836 & $<$70 & $\cdots$ & $<$1150 & $<$74 & $\cdots$ & $\cdots$ & $\cdots$ & $\cdots$ & $\cdots$ & $\cdots$ & $\cdots$ & $\cdots$\\
\vspace{-1ex} J1853+1303 & 95$_{-10}^{+9}$ & $<$370 & $\cdots$ & 96$_{-11}^{+9}$ & $\cdots$ & 113$_{-22}^{+18}$ & $-0.5_{-0.4}^{+0.5}$ & $\cdots$ & $\cdots$ & $\cdots$ & $\cdots$ & A (Const)\\
\vspace{-1ex} B1855+09 & 199.9$_{-2.3}^{+2.2}$ & $\cdots$ & $\cdots$ & $<$410 & 200.7$_{-2.8}^{+2.9}$ & 182.3$_{-9.0}^{+5.3}$ & 0.25$_{-0.07}^{+0.13}$ & $\cdots$ & $\cdots$ & $\cdots$ & 107.1$ \pm 2.8$,$54_{-16}^{+15}$ & C (PL)\\
\vspace{-1ex} J1903+0327 & $<$120 & $\cdots$ & $\cdots$ & $<$120 & $<$180 & 920$_{-250}^{+140}$ & $-5.1_{-0.8}^{+0.7}$ & $\cdots$ & $\cdots$ & $\cdots$ & $\cdots$ & C (PL)\\
\vspace{-1ex} J1909$-$3744 & 16.1$ \pm 0.4$ & $\cdots$ & 22.7$ \pm 1.1$ & 15.0$ \pm 0.4$ & $\cdots$ & 20.5$_{-0.6}^{+0.5}$ & $-$0.87$ \pm 0.09$ & $\cdots$ & $\cdots$ & $\cdots$ & 16.3$ \pm 0.5$,$-$28.7$_{-3.3}^{+2.9}$ & C (PL)\\
\vspace{-1ex} J1910+1256 & 140$_{-7}^{+6}$ & $\cdots$ & $\cdots$ & 137$ \pm 7$ & 171$_{-28}^{+26}$ & 76$ \pm 12$ & $1.3 \pm 0.3$ & $\cdots$ & $\cdots$ & $\cdots$ & $\cdots$ & C (PL)\\
\vspace{-1ex} J1911+1347 & 61.4$_{-3.4}^{+3.3}$ & $\cdots$ & $\cdots$ & 61.5$_{-3.2}^{+3.3}$ & $<$99 & 149$_{-18}^{+23}$ & $-2.5_{-0.5}^{+0.4}$ & $\cdots$ & $\cdots$ & $\cdots$ & $74 \pm 7$,$-222 \pm 44$ & C (PL)\\
\vspace{-1ex} J1918$-$0642 & 75$_{-16}^{+13}$ & $\cdots$ & $<$91 & 83$_{-18}^{+13}$ & $\cdots$ & 1.3$_{-0.6}^{+7.1}$ & $8.5_{-3.4}^{+1.1}$ & $\cdots$ & $\cdots$ & $\cdots$ & $\cdots$ & C (PL)\\
\vspace{-1ex} J1923+2515 & 285$ \pm 12$ & $<$310 & $\cdots$ & 290$_{-12}^{+11}$ & $\cdots$ & 310$_{-14}^{+13}$ & $-$0.36$ \pm 0.09$ & $\cdots$ & $\cdots$ & $\cdots$ & $\cdots$ & C (PL)\\
\vspace{-1ex} B1937+21 & 15.8$_{-0.1}^{+0.2}$ & $<$43 & 29.6$_{-5.3}^{+0.7}$ & 14.4$ \pm 0.2$ & 15.2$_{-7.0}^{+1.3}$ & 21.7$ \pm 0.3$ & $-$1.02$_{-0.04}^{+0.03}$ & 11.4$_{-0.5}^{+0.4}$ & $-3.0 \pm 0.2$ & 9.1$_{-0.6}^{+0.7}$ & 22.5$ \pm 0.3$,$-$36.4$ \pm 1.6$ \newline 22.7$ \pm 0.3$,$-$84.2$_{-3.5}^{+4.8}$,$219_{-18}^{+13}$ & E3 (LP3)\\
\vspace{-1ex} J1944+0907 & 426$ \pm 10$ & $<$180 & $\cdots$ & 440$_{-10}^{+9}$ & $\cdots$ & 207$_{-18}^{+17}$ & $2.1 \pm 0.2$ & $\cdots$ & $\cdots$ & $\cdots$ & $\cdots$ & C (PL)\\
\vspace{-1ex} J1946+3417$^{\rm a}$  & 262$ \pm 12$ & $\cdots$ & $\cdots$ & 262$_{-13}^{+12}$ & $\cdots$ & 440$ \pm 70$ & $-1.3 \pm 0.4$ & $\cdots$ & $\cdots$ & $\cdots$ & $\cdots$ & A (Const)\\
\vspace{-1ex} B1953+29 & 610$_{-17}^{+16}$ & 805$_{-38}^{+41}$ & $\cdots$ & 572$ \pm 20$ & $\cdots$ & 634$_{-17}^{+15}$ & $-$0.32$ \pm 0.05$ & $\cdots$ & $\cdots$ & $\cdots$ & $\cdots$ & C (PL)\\
\vspace{-1ex} J2010$-$1323 & 102$_{-19}^{+15}$ & $\cdots$ & $<$73 & 177$_{-20}^{+19}$ & $\cdots$ & 81$_{-23}^{+19}$ & $1.8_{-0.5}^{+0.6}$ & $\cdots$ & $\cdots$ & $\cdots$ & $\cdots$ & C (PL)\\
\vspace{-1ex} J2017+0603 & 32$_{-8}^{+7}$ & $\cdots$ & $\cdots$ & $<$650 & 27$_{-11}^{+7}$ & 8.4$_{-3.8}^{+4.0}$ & $3.3_{-0.6}^{+0.7}$ & $\cdots$ & $\cdots$ & $\cdots$ & $\cdots$ & C (PL)\\
\vspace{-1ex} J2033+1734 & 693$ \pm 22$ & 603$_{-36}^{+39}$ & $\cdots$ & 747$ \pm 31$ & $\cdots$ & 709$ \pm 24$ & 0.25$_{-0.06}^{+0.07}$ & $\cdots$ & $\cdots$ & $\cdots$ & $\cdots$ & C (PL)\\
\vspace{-1ex}J2043+1711 & 18.0$_{-4.4}^{+4.0}$ & $<$15 & $\cdots$ & 25.5$ \pm 3.7$ & $\cdots$ & $\cdots$ & $\cdots$ & $\cdots$ & $\cdots$ & $\cdots$ & $\cdots$ & A (Const)\vspace{2ex}\\
\vspace{-1ex} J2145$-$0750 & 338.4$_{-3.1}^{+3.0}$ & $\cdots$ & 389.7$_{-4.4}^{+5.0}$ & 244.5$_{-5.2}^{+4.4}$ & $\cdots$ & 329.3$_{-5.2}^{+3.9}$ & $-$0.76$ \pm 0.05$ & 188$_{-24}^{+19}$ & $-2.1_{-0.5}^{+0.4}$ & 130$_{-23}^{+27}$ & $\cdots$ & B ($N$-Band)\\
\vspace{-1ex} J2214+3000 & 136$ \pm 7$ & $\cdots$ & $\cdots$ & 136$ \pm 7$ & $<$200 & 109$_{-18}^{+20}$ & $0.5 \pm 0.4$ & $\cdots$ & $\cdots$ & $\cdots$ & $\cdots$ & A (Const)\\
\vspace{-1ex} J2229+2643 & 146$ \pm 10$ & $<$73 & $\cdots$ & 164$ \pm 12$ & $\cdots$ & 115$_{-15}^{+16}$ & $0.8 \pm 0.3$ & $\cdots$ & $\cdots$ & $\cdots$ & $\cdots$ & C (PL)\\
\vspace{-1ex} J2234+0611 & 27.7$_{-2.3}^{+2.4}$ & $\cdots$ & $\cdots$ & $<$140 & 27.9$_{-2.3}^{+2.1}$ & 41.9$_{-4.5}^{+4.4}$ & $-1.2 \pm 0.3$ & $\cdots$ & $\cdots$ & $\cdots$ & $40 \pm 5$,$-145_{-42}^{+38}$ & E2 (LP2)\\
\vspace{-1ex} J2234+0944 & 57.0$ \pm 3.4$ & $\cdots$ & $\cdots$ & 610$_{-210}^{+370}$ & 57.2$_{-3.4}^{+3.0}$ & 97$_{-19}^{+14}$ & $-1.4_{-0.4}^{+0.6}$ & $\cdots$ & $\cdots$ & $\cdots$ & $\cdots$ & A (Const)\\
\vspace{-1ex} J2302+4442 & 460$_{-43}^{+42}$ & $\cdots$ & 410$ \pm 60$ & 510$_{-60}^{+50}$ & $\cdots$ & 430$ \pm 50$ & $0.2 \pm 0.3$ & $\cdots$ & $\cdots$ & $\cdots$ & $\cdots$ & A (Const)\\
\vspace{-1ex} J2317+1439$^{\rm b}$  & 34.9$_{-4.1}^{+3.9}$ & {$<$28/$<$11} & $\cdots$ & 91.0$_{-4.1}^{+4.8}$ & $\cdots$ & 7.1$_{-2.7}^{+4.0}$ & $6.0 \pm 0.9$ & $\cdots$ & $\cdots$ & $\cdots$ & 46.4$_{-2.5}^{+2.7}$,$124 \pm 8$ & C (PL)\\
\vspace{-1ex} J2322+2057 & 44$_{-20}^{+16}$ & $<$67 & $\cdots$ & 55$_{-18}^{+16}$ & $\cdots$ & $\cdots$ & $\cdots$ & $\cdots$ & $\cdots$ & $\cdots$ & $\cdots$ & A (Const)\vspace{1ex}\\
\enddata
\tablenotetext{a}{\vspace{-2ex}No residuals with $S > 3$.}
\tablenotetext{b}{We list the rms jitter for the 327~MHz band followed by the 430~MHz band in the same column for Model B.}
\end{deluxetable*}

\end{longrotatetable}

%\newpage

\begin{figure}[t!]
\hspace{-5ex}
\includegraphics[width=0.58\textwidth]{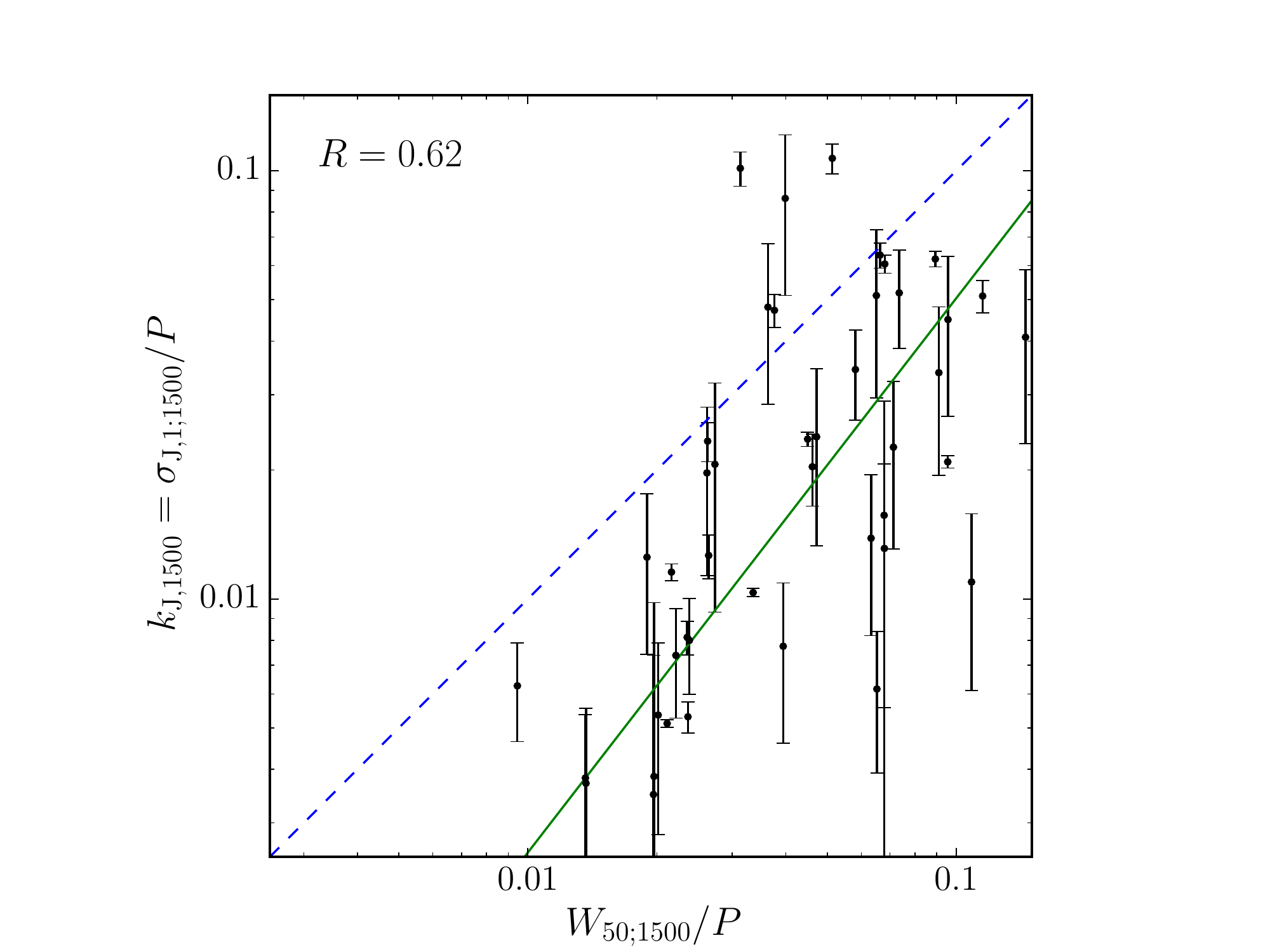}
  \caption{\footnotesize Jitter parameter $\kJref$ vs duty cycle. We use the rms jitter at 1500~MHz from the preferred model per pulsar and the full width at half maximum ($W_{50;1500}$) as a measure of the pulse width. The blue dashed line shows where both quantities are equal and the solid green line shows the best-weighted fit. The weighted correlation coefficient is $R = 0.59$.}
  \label{fig:W50corr}
\end{figure}

%We used the preferred model per pulsar and propagated uncertainties to calculate $\kJref$. We simplified the calculation by assuming all detected values were Gaussian distributed. The few upper limits on the Model A rms jitter were computed from the survival function from the Gaussian distribution with the standard deviation coming from the upper value of the confidence interval of $\sigma_{\J,1}$.

Figure~\ref{fig:W50corr} shows the correlation of $\kJref$ with the duty cycle at 1500~MHz, defined as the full width at half maximum (FWHM) of the pulse (centered around the main peak) divided by the pulse period, $W_{50;1500}/P$. While our rms jitter values represent the composite quantity for all of the individual components stochastically varying, we still expect that rms jitter should be correlated with the widths of the pulses. In nearly all cases, we see that the rms jitter is smaller the FWHM of the template profiles. Note that the FWHM may not represent the FWHM of an individual component as the ``main pulse'' (the component(s) with the highest intensity) of a pulse profile may consist of several components. Nonetheless, we find moderate correlation between $\kJref$ and the duty cycle at 1500~MHz. We calculated the square root of the coefficient of correlation as
\be 
R = \left[1 - \sum \left(\frac{\kJref - \kJhatref}{\sigma_{\kJref}}\right)^2 / \sum \left(\frac{\kJref}{\sigma_{\kJref}}\right)^2\right]^{1/2}
\ee
where $\kJhatref$ is the estimated jitter parameter from a linear fit and $\sigma_{\kJref}$ represent the errors on the measured jitter parameters. The ratio in the bracket is simply the variance of the residuals of the fit over the variance of the data. The correlation between $\kJref$ and $W_{50;1500}/P$ is $R = 0.62$.

\begin{figure}[t!]
\hspace{-5ex}
\includegraphics[width=0.58\textwidth]{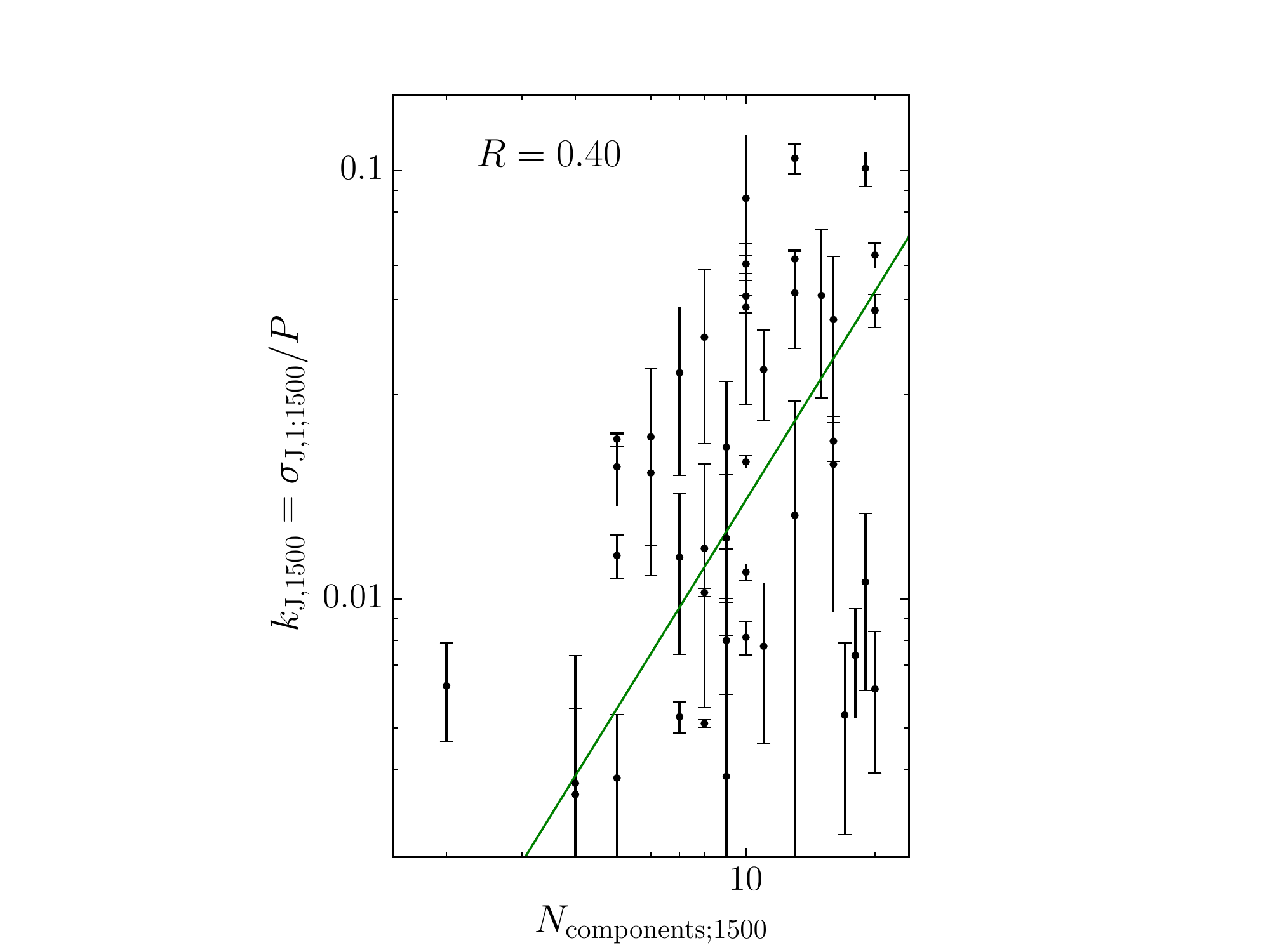}
  \caption{\footnotesize Jitter parameter $\kJref$ vs number of pulse components. The solid green line shows the best-weighted fit. The weighted correlation coefficient is $R = 0.46$.}
  \label{fig:Ncorr}
\end{figure}

Figure~\ref{fig:Ncorr} shows the correlation of $\kJref$ with the number of components $N_{\rm components;1500}$ of the pulse profile at 1500~MHz. To determine the number of components, using \pypulse we fit von Mises functions to the templates,
\be 
f(x|a,b) = \frac{e^{b\cos(x-a)}}{2 \pi I_0(b)},
\ee
the circular analog of the Gaussian function where $a$ and $b$ are the two parameters describing each von Mises function and $I_0$ is the modified Bessel function of the first kind. We performed the fit iteratively until either (i) the residuals of the templates minus the fit were less than 0.01 of the maximum, (ii) adding more components were deemed insignificant via an F-test with a significance value of 0.05 (i.e., at the ``$2\sigma$'' level), or (iii) 20 components were fit. With high S/N observations of the NANOGrav pulse profiles, \citet{Gentile+2018} have detected microcomponents with low amplitudes and thus even with a cutoff of 1\% of the maximum there may be additional observable emission for some pulsars. In this analysis, we see again moderate (but less than the duty cycle) correlation between $\kJref$ and $N_{\rm components;1500}$, suggesting that an increased number of components may lead to an increase in the amount of the total rms jitter. Note that our estimates of the number of components may be biased by the use of von Mises functions in the fit if the true emission beams are not well-described by such functional forms.

 \todoblank{theoretical basis for this}

We looked at correlations of $\kJref$ with $\Weffref$ to see if the rms jitter values were related to the ``sharpness'' of the templates (see Eq.~\ref{eq:Weff}) but found no significant relationship. We also looked for correlations between $\kJref$ and DM to see if scintillation noise misestimates were significantly biasing our results but found no relationship as well.

In addition to looking at the individual $\kJref$ values, we fit a likelihood for a single global frequency dependence of $\kJ$. We assumed a power-law form $\kJ(\nu) = \kJo (\nu/\mathrm{1000~MHz})^\gamma$ since many of the pulsars that show frequency dependence of the rms jitter have Model C as the preferred model. While variations in the Model C $\alpha$ terms are large, this modeling allows us to come up with an average frequency dependence for the rms jitter across all of our pulsars by leveraging all of the residuals simultaneously, potentially more sensitive than creating and comparing continuous histograms as in Figure.~\ref{fig:kJ}.

With the above motivation, we define the global likelihood over pulsars as 
\ba
& & \Like_{\rm global}(\kJo,\gamma | \left\{S_{n,i},\R_{n,i},\nu_{n,i},\chivec_n\right\}) = \nonumber \\
& & \prod_n \prod_i \Like(\kJo,\gamma| \left\{S_{n,i},\R_{n,i},\nu_{n,i}\right\},\chivec_n)
\label{eq:globallikelihood}
\ea
where $n$ is an index over all pulsars. As before, we used the Gaussian PDF in Eq.~\ref{eq:gaussian} to model the individual residuals and then jitter component of the total rms white noise per pulsar via Eq.~\ref{eq:sigmaJ} is simply
\be 
\sigma_\J(\nu|\kJo,\gamma) = \frac{\kJo (\nu/\mathrm{1000~MHz})^\gamma P}{\sqrt{\Np}}.
\label{eq:sigmaJ2}
\ee
After running our MCMC analysis, we estimated the global parameters to be $\kJo = 8.43 \pm 0.03 \times 10^{-3}$, $\gamma = -0.42 \pm 0.01$. When scaled to 1500~MHz with the uncertainties propagated, we find $\kJref = 7.10 \pm 0.04 \times 10^{-3}$, i.e., $\approx$0.7\% which is about half the estimate from the continuous histogram (Figure~\ref{fig:kJ}) but still consistent given the large uncertainties from that method. The error we report here is derived from the confidence intervals for the two parameters and does not include the systematic variations between the per-pulsar parameters, i.e., a wide range of estimates of $\alpha$.%values as seen with estimates of $\alpha$.

%\begin{figure}[t!]
%\hspace{-5ex}
%\includegraphics[width=0.58\textwidth]{freqdependencies.png}
%  \caption{\footnotesize Placeholder plot for frequency %dependencies? Or plot of $\alpha$ values?}
%\end{figure}

%\subsection{Correlations}
%\label{sec:correlations}

%\begin{deluxetable*}{lcccccccc}
%\tablecolumns{9}
%\tablecaption{Summary of White Noise Contributions Scaled to 30 Minutes}
%\tablehead{
%\colhead{Pulsar} & \colhead{S/N error} & \colhead{Jitter error} & \colhead{DISS error} & \colhead{Total error} &  \colhead{Rank S/N} & \colhead{Rank Jitter} & \colhead{Rank DISS} & \colhead{Rank Total} 
%}
%\startdata
%\enddata
%\end{deluxetable*}

\subsection{Interpulse Statistics}
\label{sec:interpulse}

While all of our pulsars show complicated profile shapes, several show distinct interpulses $\sim$50\% of pulse phase away from the main pulse. The only pulsar with high enough S/N in both the main pulse and interpulse for us to analyze was PSR~B1937+21. Since we observe this pulsar with both AO and GBO, we have data covering four frequency bands. The main pulse at 1400~MHz and 2300~MHz shows a second smaller component overlapping the primary in phase; the interpulse shows a similar feature. At 430~MHz and 820~MHz, pulse broadening from interstellar scattering causes the smaller components to blend into the primary ones.

We split each profile $\approx$25\% of pulse phase after the main pulse and repeated the likelihood analysis on each section individually using the preferred Model E3. While there were significant deviations in the third coefficient, $\sigma_{\J,1,2}$ (the coefficient on the quadratic term), the slight but consistent variations in the other two coefficients between the main pulse and interpulse meant that the total $\sigma_{\J,1}$ was consistent between both components and therefore we find no significiant variation between the rms jitter for both the main pulse and interpulse of PSR~B1937+21. Even though the scintillation parameters for this pulsar, and therefore $\sigma_{\DISS}$, may be changing over time, the scintillation noise should be equal for the main pulse and interpulse. Therefore, the measure of jitter between both sets will be unbiased by uncertain interstellar parameters.

When we tested the other jitter models, we found $\sim2\sigma$ variations in Model A (constant), due largely to the significant variations in the 2300~MHz band (main pulse: $\sigma_{J,1;2300} = 11.6 \pm 0.8~\mathrm{\mu s}$, interpulse: $\sigma_{J,1;2300} = 17.4 \pm 1.5~\mathrm{\mu s}$) in Model B ($N$-band); the other models were largely consistent between main pulse and interpulse. The deviation in the 2300~MHz jitter estimates causes the same shift in the Model E3 coefficient described above and so we do not believe the difference in values is intrinsic to the pulsar. The likely cause for the discrepancy is a lower interpulse S/N at 2300~MHz coupled with known RFI contamination in that band at AO.

% We find no significant variation between the rms jitter parameters for all models and for both the main pulse and interpulse of PSR~B1937+21. Even though the scintillation parameters for this pulsar, and therefore $\sigma_{\DISS}$, may be changing over time, the scintillation noise should be equal for the main pulse and interpulse. Therefore, the measure of jitter between both sets will be unbiased by uncertain interstellar parameters.

\todoblank{Number of components versus jitter, as in NG9WN Eq. 13?}

\subsection{Testing the Time-dependence of Jitter}

Just as pulse profiles are known to be stable over long periods of time, we expect that the statistics of jitter will remain stable over these timescales. For PSRs~J1713+0747 and J1909$-$3744, we looked at the time-dependence of our rms jitter measurements. We chose these two as they are high-S/N pulsars with low scintillation noise and so variations in the scintillation parameters will not bias $\sigma_\R$ largely. We repeated our main procedures of testing the various models for frequency dependence except we looked at subsets of the residuals separately in one-year bins spanning from 2010 to 2016 (inclusive, covering the total time range of our observations). 

\begin{figure}[t!]
\hspace{-7ex}
\includegraphics[width=0.55\textwidth]{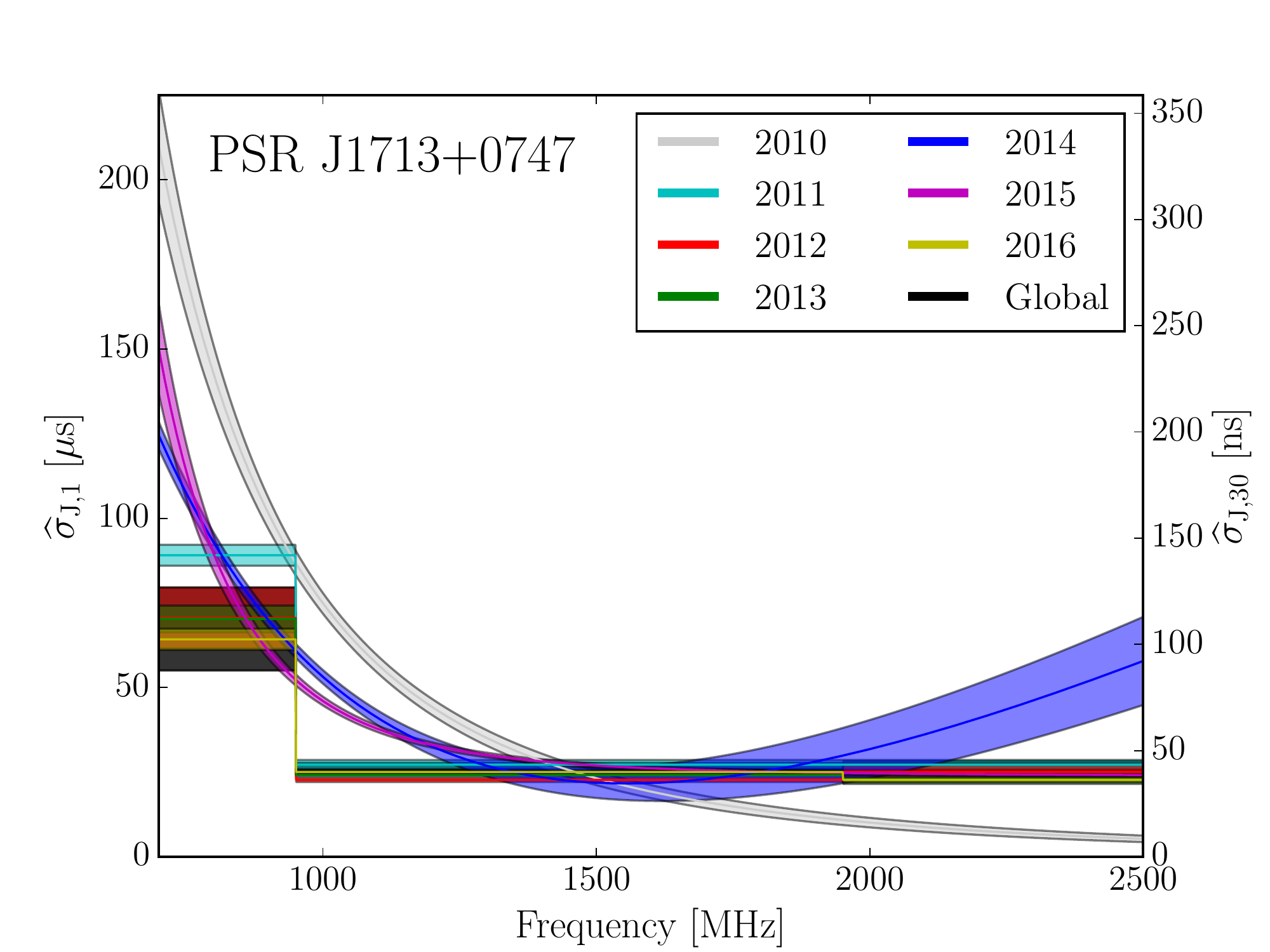}
  \caption{\footnotesize Estimated frequency-dependent jitter models per year for PSR~J1713+0747. The preferred models per year are plotted along with the global fit. The shaded regions show the $\sim1\sigma$ errors. The right-hand axis shows the 30-minute rms jitter.}
  \label{fig:timedepJ1713}
\end{figure}

Figure~\ref{fig:timedepJ1713} shows the results for PSR~J1713+0747. The breaking points for the Model B curves were chosen to be in between the edges of the frequency bands. We see fairly good agreement in the frequency-dependence of $\sigma_{\J,1}$ over time and compared to the global fit (black, drawn behind because there is a large spread over the 820~MHz band); the shaded regions show the $\sim1\sigma$ errors. However, we see significant variations in the 820~MHz band, possibly due to longer-term variations in the scintillation parameters over time resulting in biases in our jitter estimates.

% In 2010, GUPPI but not PUPPI was online and we only have data for the 820 and 1400~MHz bands; in the plots we extend the value for $\sigma_{\J,1;1400}$ up to the highest frequencies. In 2016, Model D is the second-most-significant model. Nevertheless, we see fairly good agreement in the frequency-dependence of $\sigma_{\J,1}$ over time and compared to the global fit; the shaded regions show the $\sim1\sigma$ errors.

% The two models that were significant were either Model B ($N$-Band) or Model D (Power Law plus Constant).

\begin{figure}[t!]
\hspace{-4ex}
\includegraphics[width=0.55\textwidth]{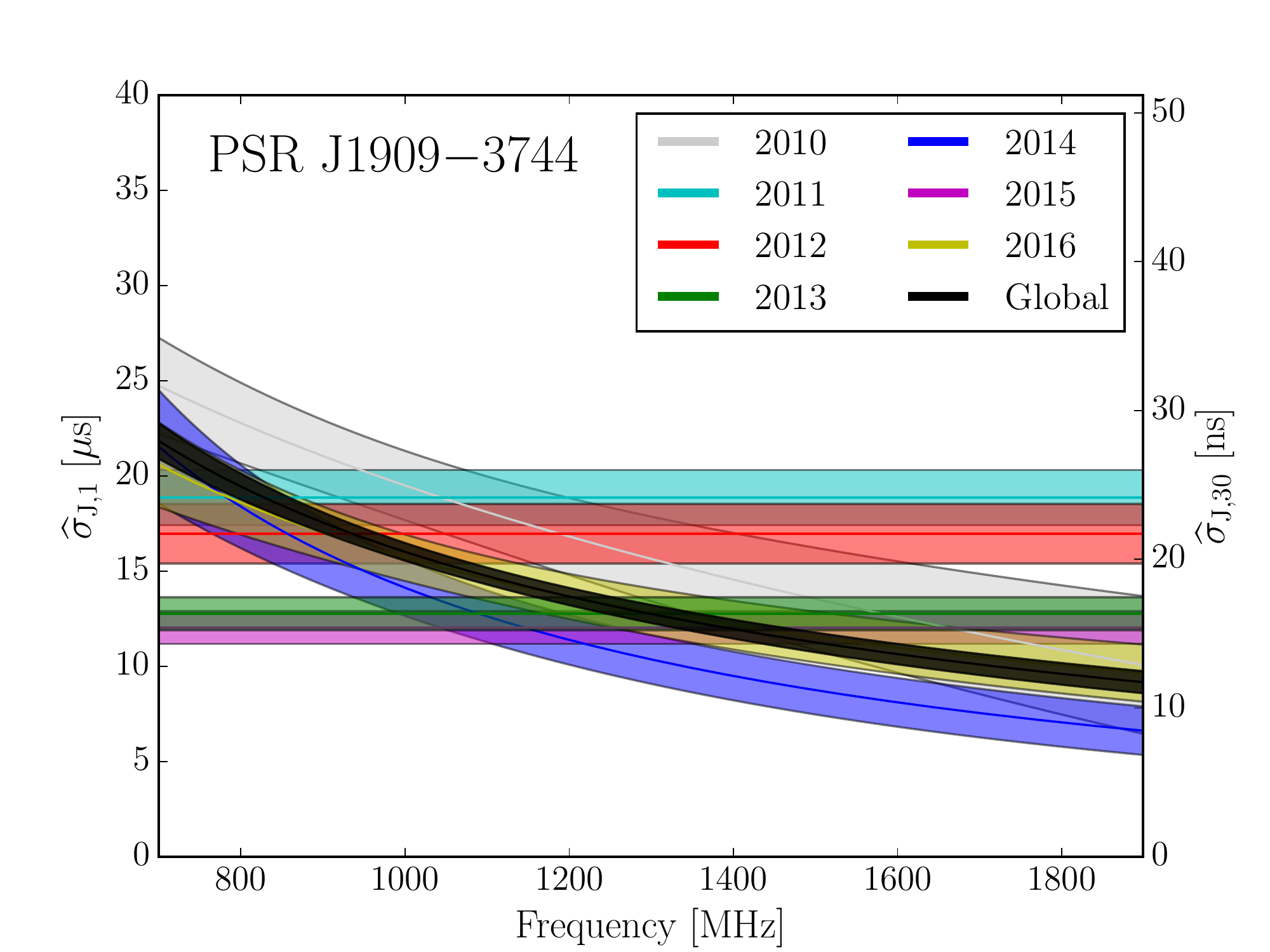}
  \caption{\footnotesize Estimated frequency-dependent jitter models per year for PSR~J1909$-$3744. The preferred models per year are plotted along with the global fit. The shaded regions show the $\sim1\sigma$ errors. The right-hand axis shows the 30-minute rms jitter. We note that our models likely overestimate the rms jitter values intrinsic to the pulsar due to contamination from other systematics as discussed in the text.}
  \label{fig:timedepJ1909}
\end{figure}

Figure~\ref{fig:timedepJ1909} shows the results for PSR~J1909$-$3744. We see significant deviations per year for the four years where Model B is significant, with the lowest values in 2013 and 2015 versus 2011 and 2012. The highest S/N data (counting all data with $S \gtrsim 100$) is in 2014; therefore one likely explanation for the higher $\sigma_{\J,1}$ values across the 1500~MHz band in other years may be due to other systematics such as RFI. Via simulations of residuals containing only template-fitting and jitter errors, we found that only a few number of high-S/N residuals (of order a few tens) were required to make a confident and unbiased detection of jitter. 

% Except for 2016 where Model E2 (Log-Polynomial with two coefficients), Model A was the most significant model per year. Recall that Model E2 was the best-fit global model.

\citet{Brook+} looked at pulse-profile variability in the NANOGrav 11-yr data set (last observation at the end of 2015) but found no significant variations for either pulsar. \citet{Levin+2016} showed no significant variations in $\taud$ in the NANOGrav 9-yr data set (last observation at the end of 2013) for PSR~J1909$-$3744 and only slight variations in PSR~J1713+0747 though with neither a long-term trend nor correlated with DM.

Long-term variations in the scinillation parameters could explain the offsets in $\sigma_{\J,1}$ for PSR~J1909$-$3744, coupled with the lack of sensitivity of our method to distinguish the frequency dependence. The upper bound on $\sigma_{\DISS}$ can be taken as $\approx \taud$ (when $\niss\to1$), and of we consider only the values across the 1500~MHz band where $\taudo = 4.1$~ns (Table~\ref{table:input}), then variations in the parameters of order a few could explain the differences in the estimated rms jitter. Therefore, we cannot say definitively if the rms jitter in PSR~J1909$-$3744 is changing over time or if biases such as from long-term variations in the scintillation parameters, or even RFI, are causing the difference we observe. Given our timing sensitivity, even subtle levels of RFI will easily affect the data; note that the 30-minute rms jitter values for PSR~J1909$-$3744 are a factor of $\sim2$ lower than for PSR~J1713+0747 at the highest frequencies but up to an order of magnitude lower than at the lowest frequencies and therefore are more susceptible to low-level systematic errors. If we assume that the 2014 value is the true value of $\sigma_{\J,1}$ and the others are biased by RFI or polarization-calibration errors, or if the scintillation parameters and thus $\sigma_{\DISS}$ values are varying over long timescales, then our estimates of the overall rms jitter intrinsic to PSR~J1909$-$3744 (again denoted with a caret) may be overestimated in Tables~\ref{table:resultsA} and \ref{table:resultsB} by a factor of $\sim$2 at least, suggesting that PSR~J1909$-$3744 may be more intrinsically higher-precision than we predict here.

\section{Discussion: The Optimization of Pulsar Timing Arrays}
\label{sec:discussion}

We have characterized the frequency dependence of jitter in pulsars observed by NANOGrav. Jitter values can be obtained robustly from a single or few high-S/N observations, so viewing a pulsar during one or a few scintillation maxima (from long integrations and/or large bandwidths) is critical. Improvements in those estimates of jitter and the frequency dependence require as many high S/N observations as possible. Varying scintillation-noise parameters bias $\sigma_{\DISS}$ and therefore our estimates of $\sigma_\J$; these parameters need to be monitored as frequently as possible for proper modeling. Scintillation parameter variations along with other interstellar propagation-delay variations will also affect the noise modeling on longer timescales (e.g., scatter broadening changes causing frequency-dependent red noise in TOAs; \citealt{Lentati+2017}).

As larger telescopes and/or telescopes with wideband receivers come online, $\sigma_{\SN}$ is becoming negligible in comparison to $\sigma_\J$ and $\sigma_{\DISS}$; already we are seeing this is true for many NANOGrav observations since we can detect jitter. For interferometers with the capability to form sub-arrays, observations can be optimized by pointing at multiple high-S/N pulsars simultaneously, allowing for an increase in the number of pulses $\Np$ and reducing the overall jitter error in favor of increasing the sub-dominant $\sigma_{\SN}$ error.

Jitter becomes dominant depending on its frequency dependence \citep{optimalfreq}. Consider jitter with a power-law scaling as in Model C in the high-S/N regime. At high frequencies, $\niss \to 1$ so $\sigma_{\DISS} \approx \taud \propto \nu^{-22/5}$. We find that $\sigma_\J$ typically has a shallower scaling than this (including for the other frequency-dependence models) and therefore $\sigma_{\DISS}$ will become negligible at the highest frequencies compared to $\sigma_\J$ as one would naively expect from observations of scatter broadening and dynamic spectra. At lower frequencies where scatter broadening increases as does the number of scintles, then $\sigma_{\DISS} \propto \nu^{-8/5}$. The cross-over frequency is then
\be 
\nu_{\rm cross} = \nuo \left(\frac{\sigma_{\J,0}}{\sigma_{\DISS,0}}\right)^{-1/(\alpha+8/5)}
\ee
where again the subscript `0' denotes quantities at some reference frequency $\nuo$. If the ratio of $\sigma_{\J,0}/\sigma_{\DISS,0}$ at 1500~MHz is roughly 10$-$100 (NG9WN) and $\alpha > -8/5$, then $\nu_{\rm cross}$ will be $<$1500~MHz, which is typically what we find. As we expect, jitter will dominate for most of our higher frequencies except for pulsars at high DM.

Similarly, in the moderate-S/N regime when we must consider template-fitting errors, i.e., when $\sigma_{\SN} \gtrsim \sigma_\J$, we have to first order $S \propto \nu^{-\alpha_{\rm flux}}$ where $\alpha_{\rm flux} \approx 1.6$ is the pulsar's flux spectral index \citep{Jankowski+2018}, and therefore $\sigma_{\SN} \propto \nu^{\alpha_{\rm flux}}$, i.e., the template-fitting errors increase at higher frequencies. This simplification ignores frequency-dependent pulse-profile evolution, pulse broadening from scattering, system temperature variations, etc. Following this simplification, as above we have
\be 
\nu_{\rm cross} = \nuo \left(\frac{\sigma_{\J,0}}{\sigma_{\SN,0}}\right)^{-1/(\alpha+\alpha_{\rm flux})}
\ee
so that since typically we have $\alpha \gtrsim -1.6$, $\nu_{\rm cross}$ will eventually dominate at higher frequencies. Therefore, for optimal precision timing with higher center frequencies, we expect that jitter will begin to become the dominant white-noise term and understanding the statistics of jitter within our data will become a requirement in the era of low-frequency GW characterization.

% since we do not see those values of $\alpha$ typically

% especially for newer telescopes with higher gains \citep{optimalfreq}. 

\acknowledgments

{\it Author contributions.} MTL developed the mathematical framework for this work, created the modified set of profiles and short-term residuals, performed the analysis, and prepared the majority of the manuscript. MAM assisted with the preparation of the manuscript. All authors performed observations and/or helped produce the initial pulse profile data set. Additional specific contributions will be described in the future NANOGrav 12.5-year data set paper.

%ZA, HB, PRB, HTC, PBD, MED, TD, JAE, RDF, ECF, EF, NGD, PAG, MLJ, DRL, RSL, CN, DJN, TTP, SMR, RS, IHS, KS, JKS, SJV, and WWZ 

% JMC, SC, GJ, and DJN helped with review of the manuscript.

{\it Acknowledgments.} The NANOGrav Project receives support from NSF Physics Frontiers Center award number 1430284. Pulsar research at UBC is supported by an NSERC Discovery Grant and by the Canadian Institute for Advanced Research. The Arecibo Observatory is operated by SRI International under a cooperative agreement with the NSF (AST-1100968), and in alliance with Ana G. M\'{e}ndez-Universidad Metropolitana, and the Universities Space Research Association. The National Radio Astronomy Observatory and the Green Bank Observatory are facilities of the National Science Foundation operated under cooperative agreement by Associated Universities, Inc.  WWZ is supported by the CAS Pioneer Hundred Talents Program and the Strategic Priority Research Program of the Chinese Academy of Sciences, Grant No. XDB23000000. 

% Part of this research was carried out at the Jet Propulsion Laboratory, California Institute of Technology, under a contract with the National Aeronautics and Space Administration.

%

\software{PSRCHIVE \citep{psrfits,psrchive}, nanopipe \citep{nanopipe}, PyPulse \citep{pypulse}}

\end{document}